\documentclass[%
 reprint,
 amsmath,amssymb,
 aps,
 nofootinbib
]{revtex4-2}

\usepackage{graphicx}
\usepackage{bm}
\usepackage[T1]{fontenc}    
\usepackage{hyperref}       
\usepackage{url}           
\usepackage{booktabs}      
\usepackage{amsfonts}      
\usepackage{nicefrac}       
\usepackage{microtype}      
\usepackage{subfig}
\usepackage[rightcaption]{sidecap}
\usepackage{wrapfig}
\usepackage{amsmath}
\usepackage{fancyvrb}
\usepackage{braket}
\usepackage{bbold}
\usepackage[table]{xcolor}
\usepackage{listings}
\usepackage{color}

\usepackage{multirow} 

\definecolor{dkgreen}{rgb}{0,0.6,0}
\definecolor{gray}{rgb}{0.5,0.5,0.5}
\definecolor{mauve}{rgb}{0.58,0,0.82}

\renewcommand{\Re}{\textrm{Re}}
\renewcommand{\Im}{\textrm{Im}}

\begin{document}

\title{Restricted Boltzmann Machine representation for the groundstate and excited states of Kitaev Honeycomb model}
\thanks{Corresponding author's email: \texttt{babakhaghighat@tsinghua.edu.cn}}%

\author{
    Mohammadreza Noormandipour\hspace{1mm}$^{a}$, Sun Youran\hspace{1mm}$^{b}$, Babak Haghighat\hspace{1mm}$^{b}$  \\
	$^{a}$ \textit{TCM Group, Cavendish Laboratory, University of Cambridge,
J.J. Thomson Avenue, Cambridge CB3 0HE, United Kingdom}\\
	$^{b}$ \textit{Yau Mathematical Sciences Center, Tsinghua University, Beijing, 100084, China}
}

\date{\today}

\begin{abstract}
In this work, the capability of restricted Boltzmann machines (RBMs) to find solutions for the Kitaev honeycomb model with periodic boundary conditions is investigated. The measured groundstate (GS) energy of the system is compared and, for small lattice sizes (e.g. $3 \times 3$ with $18$ spinors), shown to agree with the analytically derived value of the energy up to a deviation of $0.09\%$. Moreover, the wave-functions we find have $99.89\%$ overlap with the exact ground state wave-functions. Furthermore, the possibility of realizing anyons in the RBM is discussed and an algorithm is given to build these anyonic excitations and braid them for possible future applications in quantum computation. Using the correspondence between topological field theories in (2+1)d and 2d CFTs, we propose an identification between our RBM states with the Moore-Read state and conformal blocks of the $2$d Ising model.
\end{abstract}


\maketitle

\section{Introduction}

\begin{figure}[!htb]
    \centering
    \subfloat[]{\includegraphics[width=0.45\textwidth]{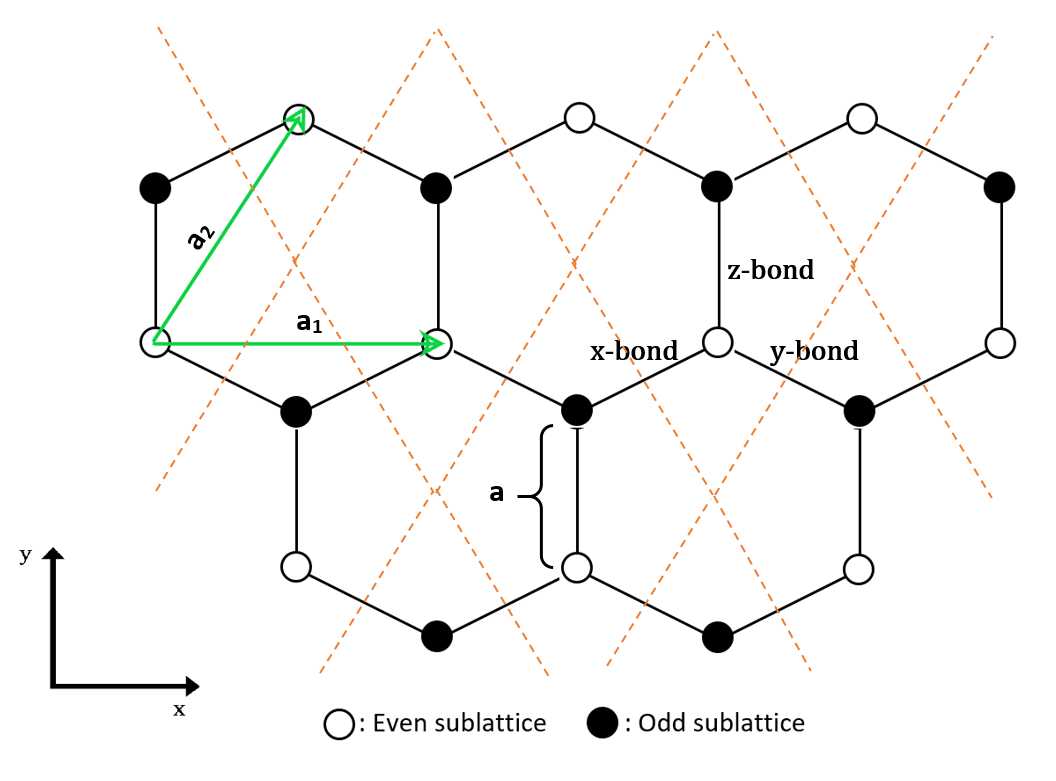}}\\
    \subfloat[]{\includegraphics[width=0.35\textwidth]{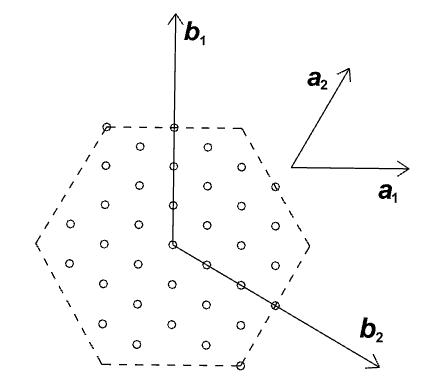}}
    \caption{(a) The honeycomb lattice in position space and its primitive vectors and primitive cell. (b) First Brillouin zone and primitive vectors of reciprocal lattice.}
    \label{fig:fig0}
\end{figure}

The Honeycomb model we consider in this work is a 2-dimensional lattice spin system (see Fig. \ref{fig:fig0}) which was studied in detail by Kitaev in 2006 \cite{Kitaev_2006} and is famous because of the topological quantum order due to a degenerate gapped groundstate which is persistent to local and finite-sized perturbations. The system also supports both Abelian and non-Abelian topological phases as demonstrated in the original proposal \cite{Kitaev_2006}. There is a wide range of applications for this model, from fault-tolerant quantum computation \cite{Kitaev_2003} to analytical study of strongly correlated systems \cite{Jackeli_2009} and quantum spin liquids \cite{Tikhonov_2011}. 

The honeycomb lattice is not a Bravias lattice in its original structure, but it can be considered as a triangular Bravias lattice with a two-spin basis. The direct Bravias lattice and the primitive cells are illustrated in Fig.  \ref{fig:fig0}a with the dashed lines. Each primitive cell has a pair of odd and even indexed spins. Just for the sake of simplicity, we will denote each cell with the even indexed (empty circles) spins of the system. The primitive vectors of the lattice, $\textbf{a}_1$ and $\textbf{a}_2$ are also shown in Fig. \ref{fig:fig0}a (see also Eq.  \eqref{eq:0}). The entire lattice can be tiled and covered with primitive cells using translations composed of different linear combinations of primitive vectors
\begin{equation}\label{eq:0}
    \textbf{a}_1 = \sqrt{3} a \textbf{e}_x \quad \& \quad
    \textbf{a}_2 = \frac{\sqrt{3}}{2} a (\textbf{e}_x,\sqrt{3}\textbf{e}_y),
\end{equation}
where $a$ is the lattice constant. Basis vectors $\textbf{b}_1$ and $\textbf{b}_2$ for the reciprocal lattice can be obtained by solving Eq.  \eqref{eq:01}
\begin{equation}\label{eq:01}
    \textbf{a}_i.\textbf{b}_j = 2\pi\delta_{ij}.
\end{equation}
The solution is as below
\begin{equation}\label{eq:02}
    \textbf{b}_1 = \frac{2\pi}{\sqrt{3}a} (\textbf{e}_x-\frac{1}{\sqrt{3}}\textbf{e}_y) \quad \& \quad
    \textbf{b}_2 = \frac{4\pi}{3a} \textbf{e}_y.
\end{equation}
The reciprocal lattice is depicted in Fig.  \ref{fig:fig0}b.

The Hamiltonian of the Kitaev model is a nearest-neighbor interaction of Pauli matrices on a honeycomb lattice as written in Eq.  \eqref{eq:1}, where $r \& r'$ are indices for the nearest neighbour spins. The physics of the system is symmetric under permutation of coupling strengths $J_{\alpha}$ with $\alpha = x, y, z$ and due to an-isotropic interaction, the model is a frustrated spin system, because a spin cannot satisfy conflicting demands of orientation from its three neighboring sites \cite{Kitaev_2006}
\begin{equation}\label{eq:1}
    H = - \sum_{\alpha} J_{\alpha} \sum_{\alpha-bonds} \sigma^{\alpha}_{r}\sigma^{\alpha}_{r'}.
\end{equation}

\begin{figure}[!htb]
	\centering
	\includegraphics[width=0.45\textwidth]{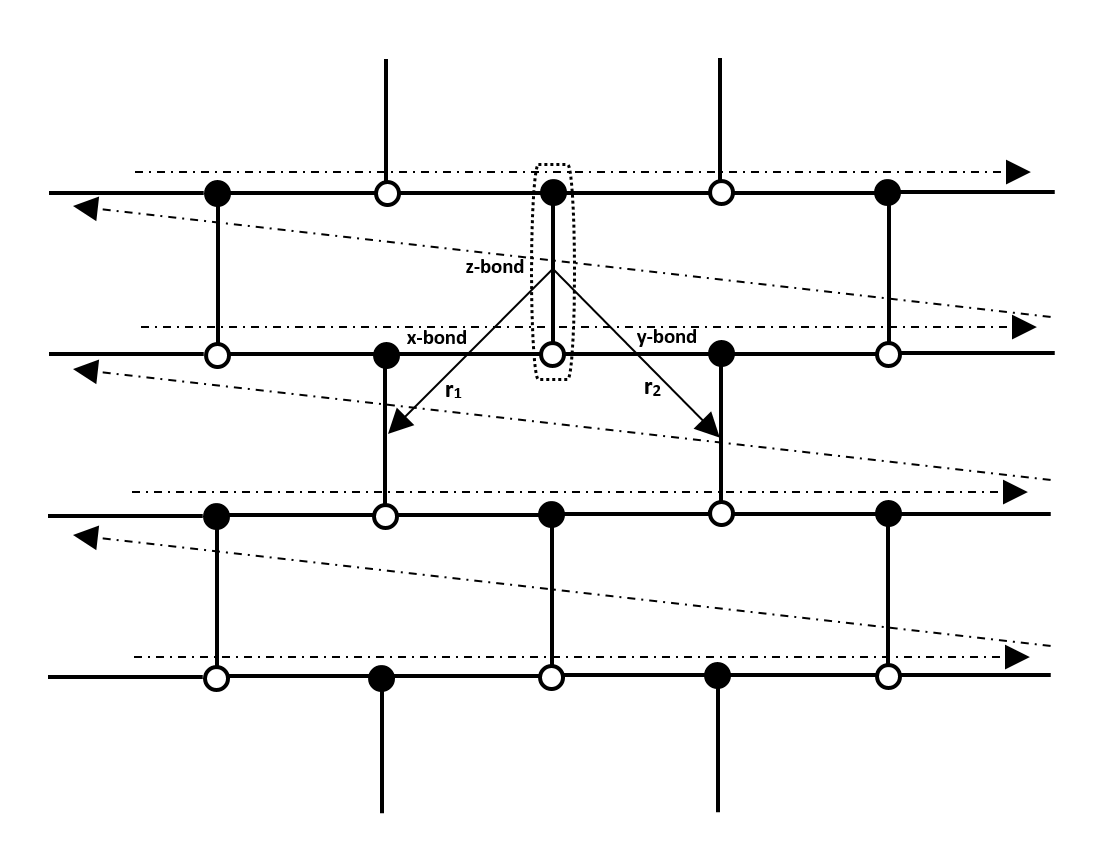}
	\caption{\label{jor} Brick-wall lattice and the Jordan-Wigner transformation path depicted with the dashed arrows. The $\textbf{r}_1$ and $\textbf{r}_2$ are the primitive vectors for this lattice.} 
	\label{fig:fig01}
\end{figure}

\begin{figure}[!htb]
	\centering
	\includegraphics[width=0.45\textwidth]{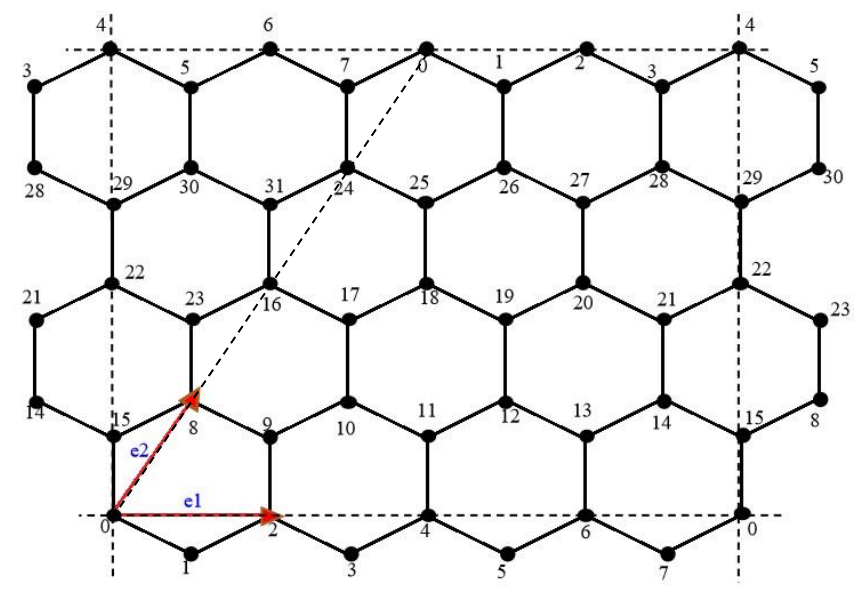}
	\caption{\label{kit_torus} Kitaev Honeycomb lattice model on a torus: The generators of the lattice in position space are named
	$\textbf{e}_{1}$ and  $\textbf{e}_{2}$, as shown in the image above. Hence, the lattice points are $\textbf{L} = n_{1}\textbf{e}_{1} + 
	n_{2}\textbf{e}_{2}$ with $n_{1}, n_{2} \in \mathbb{Z}$ and the above lattice is drawn for $N_{1} = N_{2} = 4 $ with toric boundary conditions.} 
	\label{fig:fig1}
\end{figure}

Of central interest in this work is the property of the Kitaev model to support non-Abelian chiral spin liquid (CSL) states. A very symmetric way to realize this phase of the model is by setting $J_x = J_y = J_z = 1$ and turning on a small magnetic field which gives rise to a further interaction term in the Hamiltonian (see Eq. \eqref{eq:Bfield}). In the present work, we will restrict ourselves to this non-Abelian phase but set - with the exception of Section \ref{sec:nonzeroK} - the magnetic field to zero. When CSL systems are placed on topologically non-trivial Riemann surfaces with non-zero genus, one obtains multiple degenerate groundstates that are separated from the rest of the spectrum by an energy gap. This allows to describe the low-energy physics of such models using topological quantum field theories (TQFTs) (see for example \cite{Nayak_2008} for a review). The non-Abelian phase of the Kitaev model is in the universality class of the Ising TQFT. When placed on the torus, which corresponds to periodic boundary conditions of the lattice and is the choice employed in this work, this results in a triply-degenerate groundstate. These $3$ groundstates correspond to the $3$ different ways to obtain a torus by gluing the two ends of a cylinder in the Ising TQFT, namely by inserting the operators $\mathbb{1}$, $\sigma$ and $\psi$ at the two ends.

For the topological phase discussed above, the Kitaev model has recently been numerically solved in the spin basis to a high degree of approximation using tensor networks \cite{Lee_2019,Lee_2020}. In this paper we employ a different method based on the approach developed in \cite{Deng_2017} using RBMs to find groundstates and excited states of the Honeycomb model. For a different application of RBMs to the Kitaev model, see \cite{Alcalde_Puente_2020}. The RBM as a variational ansatz to find groundstates of spin lattice models was first proposed in \cite{Carleo_2017}, where the formalism is applied to the 1d transverse-field Ising model and the 2d anti-ferromagnetic Heisenberg model. The authors of \cite{PhysRevLett.121.167204} use RBMs and feed-forward neural networks to study excited states of the 1d Heisenberg model and the 1d Bose-Hubbard model. Further applications of RBMs in the literature can be found in \cite{PhysRevLett.122.250501,PhysRevLett.122.250503,PhysRevLett.122.250503,PhysRevB.99.214306,PhysRevA.99.012307,PhysRevB.99.155136}. An application of RBMs to systems with non-Abelian symmetries was presented in \cite{PhysRevLett.124.097201}, where the authors sample in the basis of irreducible representations instead of individual spins. 

In this work, the central focus is on the RBM as a variational ansatz to find accurate groundstate wave-functions for the $3 \times 3$ Kitaev Honeycomb lattice as this is the smallest lattice size where the model admits topological order resulting in a triply-degenerate groundstate. Along the way, we also use the techniques developed to solve accurately for groundstates of smaller lattice sizes and explore the accuracy for larger lattices as well as the possibility to realize excited states. It is, however, not the goal of this paper to develop a machine learning architecture which solves efficiently and accurately for groundstates of large lattice systems. As explained in the main body of the paper, such an endeavor would need the simultaneous deployment of several distinct techniques employed and is left for future work.

Apart from a variational ansatz to find groundstates, another deep insight into chiral topological order is provided by the work of Moore and Read \cite{Moore:1991ks}, where the authors find that many wave-functions can be expressed as conformal blocks. Conformal blocks are chiral correlation functions in certain conformal field theories (CFTs) which are also known to arise as states in the Hilbert space of the $2+1$-dimensional topological Chern-Simons theory \cite{Witten:1988hf}. This connection has been generalized to describe chiral topological states in lattice systems \cite{PhysRevLett.108.257206}. In our work, we explore the possibility to realize the $3$ groundstates of the Kitaev model on the torus in terms of conformal blocks of the $SU(2)_2$ WZW CFT and show that our ansatz respects the symmetries of the Kitaev Hamiltonian \eqref{eq:1}. The consequences of the identification of the RBM states with the conformal black ansatz is then explored in the conclusions.

The outline of this paper is as follows. In Section \ref{sec2}, we briefly review the analytical solution of the model. In Section \ref{sec3} the mapping of the spin model to an RBM architecture is explained. In particular, we demonstrate how to calculate the groundstate and excited states of the Honeycomb model using machine learning techniques. To this end we employ both the \texttt{NetKet} environment \cite{netket:2019} as well our own self-built package using PyTorch to train a restricted Boltzmann machine (RBM) in order to find a groundstate via gradient descent. In Section \ref{sec4} realization of vortices is discussed from both analytical and RBM points of view. In brief, a vortex pair can be created by adjusting the parameters of the RBM and subsequent braiding can be achieved by further adjustments of parameters. Section \ref{sec5} is dedicated to results and relevant discussions and the reconstruction of a given quantum state from simple measurements known as quantum state tomography \cite{Torlai_2018}. In Section \ref{sec:cft} we propose an identification of our RBM states with the Moore-Read states on the torus which can be computed using conformal blocks of the 2d Ising model. Finally we conclude in Section \ref{sec6}.

\section{Solution of the Model}\label{sec2}

In this section an exact solution of the system is provided and an explicit form of the groundstate energy is derived where we will be following references \cite{Kitaev_2006,pachos_2012,Schmoll_2017}. First of all, we fermionize the Hamiltonian by performing a one-dimensional Jordan-Wigner transformation \cite{Jordan:1928wi} defined in Eq.  \eqref{eq:2}. A deformation of the hexagonal lattice to a brick-wall lattice (see Fig. \ref{fig:fig01})  clarifies the mechanism of the Jordan-Wigner transformation and why it is one-dimensional \cite{Feng_2007,Chen_2008}. In the brick-wall lattice, each lattice site $r$ is denoted by the coordinates $(i,j)$.
\begin{equation}\label{eq:2}
	\begin{aligned}
		\sigma^{+}_{i,j} &= 2[\prod_{j'<j}\prod_{i'}\sigma^{z}_{i',j'}][\prod_{i'<i}\sigma^z_{i',j}] a^{\dagger}_{i,j}\\
		\sigma^{-}_{i,j} &= 2[\prod_{j'<j}\prod_{i'}\sigma^{z}_{i',j'}][\prod_{i'<i}\sigma^z_{i',j}] a_{i,j}\\
		\sigma^{z}_{i,j} &= 2a^{\dagger}_{i,j}a_{i,j} - 1
	\end{aligned}
\end{equation}

\noindent This transformation maps the Hilbert space of spins to the Hilbert space of spinless complex fermions. Using the fact that $ \sigma^{\pm} = \sigma^{x}\pm i\sigma^{y} $ one can expand the Hamiltonian in Eq.  \eqref{eq:1} into three terms and rewrite them as below \cite{Schmoll_2017}
\begin{equation}\label{eq:3}
	\begin{aligned}
		\sigma^{x}_{i,j}\sigma^{x}_{i+1,j} &= \prod_{i'<i}\sigma^z_{i',j} (a^{\dagger}_{i,j} + a_{i,j}) \prod_{i'<i+1}\sigma^z_{i',j} (a^{\dagger}_{i+1,j} + a_{i+1,j})\\
		&= (a^{\dagger}_{i,j} + a_{i,j}) \sigma^z_{i,j} (a^{\dagger}_{i+1,j} + a_{i+1,j})\\
		&= -(a^{\dagger}_{i,j} - a_{i,j}) (a^{\dagger}_{i+1,j} + a_{i+1,j})\\
		\sigma^{y}_{i,j}\sigma^{y}_{i+1,j} &= -\prod_{i'<i-1}\sigma^z_{i',j} (a^{\dagger}_{i-1,j} - a_{i-1,j}) \prod_{i'<i}\sigma^z_{i',j} (a^{\dagger}_{i,j} - a_{i,j})\\
		&= -(a^{\dagger}_{i-1,j} - a_{i-1,j}) \sigma^z_{i-1,j} (a^{\dagger}_{i,j} - a_{i,j})\\
		&= (a^{\dagger}_{i-1,j} + a_{i-1,j}) (a^{\dagger}_{i,j} - a_{i,j})\\
		\sigma^{z}_{i,j}\sigma^{z}_{i,j+1} &=  (2a^{\dagger}_{i,j}a_{i,j} - 1) (2a^{\dagger}_{i,j+1}a_{i,j+1} - 1)
	\end{aligned}
\end{equation}

\noindent Therefore, the Hamiltonian transforms to
\begin{equation}\label{eq:4}
	\begin{aligned}
		H= &+J_x \sum_{x-links} (a^{\dagger}_{i,j} - a_{i,j}) (a^{\dagger}_{i+1,j} + a_{i+1,j})\\
		&-J_y \sum_{y-links} (a^{\dagger}_{i-1,j} + a_{i-1,j}) (a^{\dagger}_{i,j} - a_{i,j})\\
		&-J_z \sum_{z-links} (2a^{\dagger}_{i,j}a_{i,j} - 1) (2a^{\dagger}_{i,j+1}a_{i,j+1} - 1).
	\end{aligned}
\end{equation}

\noindent The $J_x$ and $J_y$ terms are quadratic interactions in spinless fermions and are easy to solve, however, the $J_z$ term is a product of number density operators and can be further simplified by introducing the Majorana operators in Eq.  \eqref{eq:5} \cite{Schmoll_2017}. As we will see, this simplification can be done due to the presence of the conserved quantity called plaquette operator $B_p$ \cite{Schmoll_2017}.
\begin{equation}\label{eq:5}
	\begin{aligned}
		c_{i,j} &= i(a^{\dagger}_{i,j} - a_{i,j}),
		    \, d_{i,j} = a^{\dagger}_{i,j} + a_{i,j}
		    \; \textrm{for} \; i+j=\textrm{even} \equiv \circ\\
		c_{i,j} &= a^{\dagger}_{i,j} + a_{i,j},
		    \, d_{i,j} = i(a^{\dagger}_{i,j} - a_{i,j})
		    \; \textrm{for} \; i+j=\textrm{odd} \equiv \color{black}\bullet
	\end{aligned}
\end{equation}

\noindent These operators have the following commutation relations
\begin{equation}\label{eq:6}
	\begin{aligned}
		&c^2_{i,j} = d^2_{i,j} = 1\\
		&\{ c_{i,j},c_{i',j'} \} = \{ d_{i,j},d_{i',j'} \} = 2\delta_{ii'} \delta_{jj'}\\
		&\{ c_{i,j},d_{i',j'} \} = 0
	\end{aligned}
\end{equation}

\noindent Then, the $J_z$ term can be rewritten using the Majorana operators
\begin{equation}\label{eq:7}
	\begin{aligned}
		\sigma^{z}_{i,j}\sigma^{z}_{i,j+1} &=  (2a^{\dagger}_{i,j}a_{i,j} - 1) (2a^{\dagger}_{i,j+1}a_{i,j+1} - 1)\\
		&= i(id_{i,j+1}d_{i,j})c_{i,j+1}c_{i,j}.
	\end{aligned}
\end{equation}

\noindent Finally, using the circle indices for the odd and even lattice sites as defined in Eq.  \eqref{eq:5}, the Hamiltonian transforms to the expression in Eq.  \eqref{eq:8}.
\begin{equation}\label{eq:8}
	\begin{aligned}
		H= &-iJ_x \sum_{x-links} c_{\circ}c_{\color{black}\bullet}\\
		&+iJ_y \sum_{y-links} c_{\color{black}\bullet}c_{\circ}\\
		&-iJ_z \sum_{z-links} (id_{\color{black}\bullet}d_{\circ})c_{\color{black}\bullet}c_{\circ}
	\end{aligned}
\end{equation}

\noindent If we write the Hamiltonian as a sum over unit cells we have
\begin{equation}\label{eq:9}
	\begin{aligned}
		H = i\sum_{\textbf{r}}&\left[J_x c_{\color{black}\bullet,\textbf{r}}c_{\circ,\textbf{r}+\textbf{r}_1} + J_y c_{\color{black}\bullet,\textbf{r}}c_{\circ,\textbf{r}+\textbf{r}_2}\right.\\
		&\left.J_z (id_{\color{black}\bullet,\textbf{r}}d_{\circ,\textbf{r}})c_{\color{black}\bullet,\textbf{r}}c_{\circ,\textbf{r}}\right],
	\end{aligned}
\end{equation}
where \textbf{r} is the position vector of z-bonds or unit cells  (see Fig.  \ref{fig:fig01}). It makes no difference in the physics if we set \textbf{r} to be any point along the z-bond, whether we choose an even site or an odd site or some point in the middle, they are all related through translation. 
What will be important in what follows, is that now since we have grouped a pair of even and odd sites as an unit cell, to evaluate the summation over the unit cells, it is sufficient to just run over the even or odd sites.

Furthermore, the $\alpha_\textbf{r} = (id_{\color{black}\bullet,\textbf{r}}d_{\circ,\textbf{r}})$ operators are defined on each z-bond of the lattice (labeled by \textbf{r}) and they commute with the Hamiltonian and are good quantum numbers.
Moreover, it can easily be shown that each plaquette operator can be written as
\begin{equation}\label{eq:10}
	B_p = \sigma^y_1\sigma^z_2\sigma^x_3\sigma^y_4\sigma^z_5\sigma^x_6 = \alpha_{61}\alpha_{43}.
\end{equation}
On the other hand, from Lieb's theorem \cite{Lieb_1994} we know that the groundstate manifold is obtained by setting $B_p = 1, \forall p$. 
Thus the uniform choice of $\alpha_\textbf{r}=1, \forall \textbf{r}$ corresponds to a vortex-free sector, nevertheless all configurations leading to the same sector are equivalent.

We also introduce a Dirac fermion on each z-link using the Majorana operators
\begin{equation}\label{eq:11}
	\begin{aligned}
		d_\textbf{r} &= \frac{1}{2} (c_{\color{black}\bullet,\textbf{r}} - ic_{\circ,\textbf{r}})\\
		d^\dagger_\textbf{r} &= \frac{1}{2} (c_{\color{black}\bullet,\textbf{r}} + ic_{\circ,\textbf{r}}).
	\end{aligned}
\end{equation}
Using the inverse transformation we can rewrite the Hamiltonian as
\begin{equation}\label{eq:12}
	\begin{aligned}
		H= \sum\nolimits_{\textbf{r}}&\Big[J_x (d^\dagger_{\textbf{r}} + d_{\textbf{r}})(d^\dagger_{\textbf{r}+\textbf{r}_1} - d_{\textbf{r}+\textbf{r}_1})\\
		&+J_y (d^\dagger_{\textbf{r}} + d_{\textbf{r}})(d^\dagger_{\textbf{r}+\textbf{r}_2} - d_{\textbf{r}+\textbf{r}_2}) \\
		&+J_z \alpha_\textbf{r}(2d^\dagger_{\textbf{r}}d_{\textbf{r}}-1)\Big].
	\end{aligned}
\end{equation}
The Hamiltonian is translation invariant and can be transformed to momentum space in order to be diagonalized. We define the Fourier transform for the Dirac fermion through
\begin{equation}\label{eq:13}
	\begin{aligned}
		d_{\textbf{r}} &= \frac{1}{\sqrt{N}}\sum_\textbf{k} e^{+i\textbf{k}.\textbf{r}}d_{\textbf{k}} \\
		d^\dagger_{\textbf{r}} &= \frac{1}{\sqrt{N}}\sum_\textbf{k} e^{-i\textbf{k}.\textbf{r}}d^\dagger_{\textbf{k}}.
	\end{aligned}
\end{equation}
Setting $\textbf{k} \rightarrow \textbf{-k}$ in $d^\dagger_\textbf{r}$ simplifies the calculations. We then have

\begin{equation}\label{eq:14}
	\begin{aligned}
		\textbf{X}: \hspace{2mm} &\frac{J_x}{N}\sum_\textbf{r}\sum_{\textbf{k},\textbf{k}'}e^{+i\textbf{k}.\textbf{r}}e^{+i\textbf{k}'.(\textbf{r}+\textbf{r}_1)}(d^\dagger_{-\textbf{k}} + d_{\textbf{k}})(d^\dagger_{-\textbf{k}'} - d_{\textbf{k}'})\\
		&= J_x\sum_{\textbf{k}} [-2\cos(k_1)d^\dagger_{\textbf{k}}d_{\textbf{k}} + i\sin(k_1)(d^\dagger_{\textbf{k}}d^\dagger_{-\textbf{k}}-h.c.)]\\
		\textbf{Y}: \hspace{2mm} &\frac{J_y}{N}\sum_\textbf{r}\sum_{\textbf{k},\textbf{k}'}e^{+i\textbf{k}.\textbf{r}}e^{+i\textbf{k}'.(\textbf{r}+\textbf{r}_2)}(d^\dagger_{-\textbf{k}} + d_{\textbf{k}})(d^\dagger_{-\textbf{k}'} - d_{\textbf{k}'})\\
		&= J_y\sum_{\textbf{k}} [-2\cos(k_2)d^\dagger_{\textbf{k}}d_{\textbf{k}} + i\sin(k_2)(d^\dagger_{\textbf{k}}d^\dagger_{-\textbf{k}}-h.c.)]\\
		\textbf{Z}: \hspace{2mm} &\frac{J_z}{N}\sum_\textbf{r}\sum_{\textbf{k},\textbf{k}'}e^{+i\textbf{k}.\textbf{r}}e^{+i\textbf{k}'.\textbf{r}}(2d^\dagger_{-\textbf{k}}d^\dagger_{\textbf{k}'}-1)\\
		&= J_z\sum_{\textbf{k}} 2d^\dagger_{\textbf{k}}d_{\textbf{k}}-J_zN.
	\end{aligned}
\end{equation}

\noindent For the above equation we used the orthogonality relation in Fourier transformation
\begin{equation}\label{eq:15}
	\sum_\textbf{r} e^{i(\textbf{k}+\textbf{k}').\textbf{r}} = N\delta(\textbf{k} + \textbf{k}').
\end{equation}

Using all previous transformations we obtain a Hamiltonian which is quadratic in the Dirac fermions living on each of the unit cells along the z-links
\begin{equation}\label{eq:16}
	\begin{aligned}
		H &= \sum_\textbf{k}[\epsilon_\textbf{k}d^\dagger_{\textbf{k}}d_{\textbf{k}} + \frac{1}{2}(i\Delta_\textbf{k}d^\dagger_{\textbf{k}}d^\dagger_{-\textbf{k}}-i\Delta_\textbf{k}d_{-\textbf{k}}d_{\textbf{k}})]-J_zN\\
		\epsilon_\textbf{k} &= 2[J_z-J_x\cos(k_1)-J_y\cos(k_2)]\\
		\Delta_\textbf{k} &= 2[J_x\sin(k_1)+J_y\sin(k_2)]
	\end{aligned}
\end{equation}
By applying a unitary Bogoliubov transformation, we diagonalize the Hamiltonian which can be written as
\begin{equation}\label{eq:17}
	\begin{aligned}
		H &= \sum_\textbf{k} \frac{1}{2} 
		\begin{pmatrix}
			\gamma^\dagger_\textbf{k} & \gamma_{\textbf{-k}}
		\end{pmatrix}
		\begin{pmatrix}
			E_\textbf{k} & 0\\
			0 & -E_\textbf{k}
		\end{pmatrix}
		\begin{pmatrix} \gamma_\textbf{k} \\ \gamma^\dagger_{\textbf{-k}} \end{pmatrix}\\
		&= \sum_\textbf{k}E_\textbf{k}(\gamma^\dagger_\textbf{k}\gamma_{\textbf{k}}-\frac{1}{2}), \quad E_{\mathbf{k}} = \sqrt{\epsilon_{\mathbf{k}}^2 + \Delta_{\mathbf{k}}^2},
	\end{aligned}
\end{equation}
where $\gamma^{\dagger}_{\mathbf{k}}$ and $\gamma_{\mathbf{k}}$ are quasiparticle creation and annihilation operators. The groundstate energy can easily be read off from the above Hamiltonian and is given by $E_{GS} = -\sum_{\mathbf{k}}\frac{E_\textbf{k}}{2}$.

\section{Restricted Boltzmann Machine Representations}\label{sec3}

\subsection{RBM as a Neural Network Quantum State}

With the ever growing applications of neural networks in sciences and the emergent new technologies to deploy and build physical neural networks, this is the right time to investigate potential applications of them in condensed matter systems. Recently, a new approach has been proposed for simulating topological quantum states using neural networks \cite{Deng_2017}. In this section we use the same approach and use a restricted Boltzmann machine (RBM) as a variational ansatz for the Honeycomb Kitaev model. There are many reasons why an RBM is chosen. This particular architecture has proved to be effective in many tasks such as dimensional reduction, classification, regression, collaborative filtering, feature learning and topic modeling and moreover a general theorem shows that RBMs are universal approximators of discrete distributions \cite{LeRouxBengio}.

\begin{figure}[!htb]
	\centering
	\includegraphics[width=0.4\textwidth]{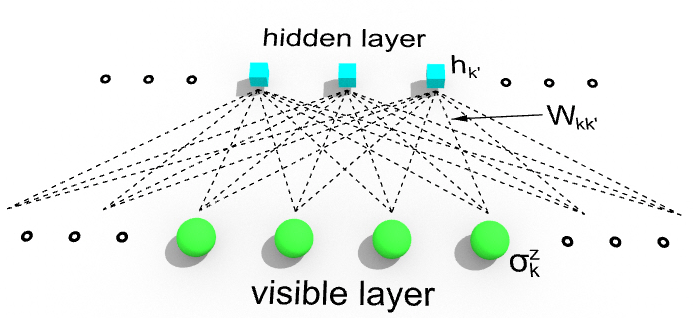}
	\caption{\label{rbm-arc} Fully connected Restricted Boltzmann machine architecture.}
\end{figure}

Having a set of spins on a lattice, $\Xi=(\sigma_{1},\sigma_{2},\cdots\sigma_{N})$ (in our case the Jordan-Wigner chain of spins), our goal is to use an RBM to reduce the dimensionality of the Hilbert space and estimate the energy of the system in both groundstate and excited states to be able to classify different topological phases of the model. An RBM consists of two layers. The first layer (visible layer) has $N$ nodes which are representing the physical spins in the Hamiltonian and the second layer (hidden layer) has $M$ binary valued nodes (architecture of the network is shown in Fig. \ref{rbm-arc}).

The quantum state of the Honeycomb model, up to an irrelevant normalization factor, can be written as 

\begin{equation}\label{eq:19_0}
    |\Phi\rangle=\sum_{\Xi}\Phi_{M}(\Xi;\Omega)|\Xi\rangle
\end{equation}
where
\begin{equation}\label{eq:19_1}
    \begin{aligned}
        \Phi_{M}(\Xi;\Omega) &=  \sum_{\{h_{k}\}}e^{\sum_{k}a_{k}\sigma_{k}^{z}+\sum_{k'}b_{k'}h_{k'}+\sum_{kk'}W_{kk'}h_{k}\sigma_{k'}^{z}}\\
        &= e^{\sum_k a_k \sigma_k^z} \times \prod_{k'} \cosh \left(\sum_k W_{kk'} \sigma_k^z + b_{k'} \right)
    \end{aligned}
\end{equation}
To obtain the second equality we used the values for $\{h_{k}\}=\{-1,1\}^{M}$ which is the set of possible configurations of hidden layer nodes and $\Omega=(a_{k},b_{k'},W_{kk'})$ is the set of weights and biases of the RBM which should be trained in such a way that the final RBM state represents the desired quantum state of the model (i.e. groundstate or the excited states), where we refer to \cite{netket:2019}, sections 2.2.3 and 2.2.4, for further details of the variational principle. The combined number of weight and bias parameters and nodes is polynomial in system size and computationally feasible.

The variational calculations for the model with RBM as ansatz is done twofold. One approach we follow is to apply the machinery already developed in the \texttt{NetKet} software package \cite{netket:2019}, while the second approach we employ is to develop our own package using PyTorch. In the case of \texttt{NetKet}, in order to train the network, the parameter space of the network is sampled using the Metropolis algorithm and the optimization iterations are based on stochastic gradient descent. It turns out that although the sampling algorithm is effective, the ground state energy cannot be reached to a satisfactory precision. The reasons will be explained as we proceed. To circumvent this problem, we develop our own software package using PyTorch which is designed tightly to suit the purposes of the Honeycomb model and avoid possible pitfalls. 

\subsection{Quantum State Tomography}
\label{sec3.2}

Reconstruction of complex synthetic quantum states from experimental measurement data is computationally exponentially expensive. Therefore, the capability of neural network quantum states such as RBM to accurately and efficiently represent high-dimensional quantum states is beneficial to quantum state tomography. Apart from the numerous applications of QST, in our study case, the reconstructed states of the honeycomb model can later on be used for validation of particular quantum processes, e.g. braiding of vortices in the system, phase transitions, etc. at the experimental level and after implementation of the framework developed in Section \ref{sec4}. For instance, there are promising proposals for implementation of topological systems in cold atoms. A review has recently been published here \cite{coldatom}. 

In Section \ref{sec5.3} we present QST results for the groundstate of a small system size of the honeycomb model, as proof of concept. In a normal QST task, the data sets come from experimental measurements, but since we don't have experimental data, we need to first find the exact wave-function for the system through exact diagonalization methods and then sample that wave-function by performing single shot measurements in different bases to build the data set. 

\section{Braiding}\label{sec4}

In this section we want to explore the possibility to create anyonic excitations in the RBM representation and describe anyon braiding. Such a representation is important for various reasons, in particular for performing quantum computation \cite{Nayak_2008}. 

\begin{figure}[!htb]
    \centering
    \includegraphics[width=0.25\textwidth]{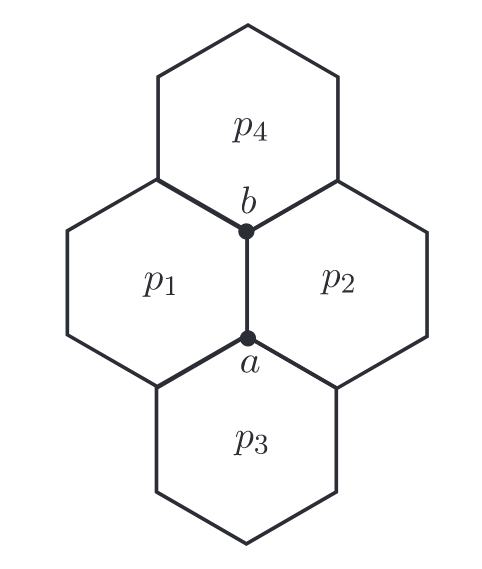}
    \caption{Demonstration of vortex pair realization using operators $\hat{O}_{1,2}$.}
    \label{fig:anyons}
\end{figure}

First of all, let us describe the creation of these quasi-particles by starting with a groundstate sector and then subsequently changing the eigenvalue of the $B_p$ operators from +1 to -1 locally and in the region where we want to realize the vortices in the system. 
In an arbitrary configuration of spins (not necessarily in the groundstate sector) it is easy to prove that $\prod B_p = +1 ,\forall p$. Hence, we can just build the vortices in pairs. For example, if you apply the operator $\hat{O}_1$ to the groundstate,  it produces two vortices in plaquettes 1 and 2 (see Fig. \ref{fig:anyons}) \cite{Leggett}.
\begin{equation}\label{eq:18}
    \hat{O}_1 = \exp{(-i\frac{\pi}{2}\hat{\sigma}^{z}_a)}
\end{equation}
Another possibility is to apply the following operator 
\begin{equation}\label{eq:19}
    \hat{O}_2 = \exp{(-i\frac{\pi}{2}\hat{\sigma}^{x}_a)}\exp{(-i\frac{\pi}{2}\hat{\sigma}^{y}_b)}~,
\end{equation}
which produces two vortices along the $\hat{z}$ direction in plaquettes 3 and 4.

{}


Let us explain how these operators create vortices from the viewpoint of Kitaev's original Majorana fermion representation \cite{Kitaev_2006}. We assume that there are two fermionic modes living on each lattice site corresponding to the four creation and annihilation operators $a^\dagger_{m,i}$ and $a_{m,i}$, where $m\in \{1,2\}$ is the index for the modes and $i$ is the index for the lattice sites. Here unlike before, we use one index for lattice site to avoid unnecessary complications. Then we will separate the imaginary and real parts of these operators, to define the Majorana fermions as below
\begin{equation}\label{eq:20}
    \begin{aligned}
        & c_i = a_{1,i} + a^\dagger_{1,i} \\
        & b^x_i = i(a^\dagger_{1,i} - a_{1,i}) \\
        & b^y_i = a_{2,i} + a^\dagger_{2,i} \\
        & b^z_i = i(a^\dagger_{2,i} - a_{2,i}).
    \end{aligned}
\end{equation}
Notice that the spins have a two dimensional space (being up or down) and now that we are representing them with four Majorana modes (two complex fermionic modes), we need to project out the unphysical states. The Fock space of the complex fermionic modes can be represented as $\{\ket{00}, \ket{01}, \ket{10}, \ket{11}\}$. We make the following correspondence between the states of spins and fermions
\begin{equation}\label{eq:21}
    \begin{aligned}
        & \ket{\uparrow} = \ket{00} \\
        & \ket{\downarrow} = \ket{11}
    \end{aligned}
\end{equation}
One can define the projector $P_i$ on site $i$ \cite{Pedrocchi_2011} to do this job for us
\begin{equation}\label{eq:22}
    \begin{aligned}
        P_i =& \frac{1+D_i}{2} \\
        D_i =& (1-2a^\dagger_{1,i}a_{1,i})(1-2a^\dagger_{2,i}a_{2,i}) \\
        =& b^x_i b^y_i b^z_i c_i~.
    \end{aligned}
\end{equation}
One can easily show that the relation between Majorana operators and the original Pauli operators is as follows
\begin{equation}\label{eq:23}
    \begin{aligned}
        \sigma^\alpha_i = i b^\alpha_i c_i \hspace{3mm}\text{for}\hspace{3mm} \alpha\in \{x,y,z\}~.\\
    \end{aligned}
\end{equation}
In fact, by the above projection, we satisfied the extra condition coming from the algebra of Pauli matrices, $-i\sigma^x_i\sigma^y_i\sigma^z_i = b^x_i b^y_i b^z_i c_i = \mathbb{1}$. Therefore, the eigenvalue of the projector $P_i$ for physical states is 1 and for unphysical states is 0.

Using Eq.  \eqref{eq:23} the Hamiltonian can be written in terms of Majorana operators as
\begin{equation}\label{eq:24}
	\begin{aligned}
		&H = \frac{i}{2} \sum_{i,j} A_{ij} c_i c_j \\ 
		&A_{ij} = J_{ij} u_{ij} \hspace{2mm}\text{,}\hspace{2mm} u_{ij} = i b^\alpha_i b^\alpha_j \hspace{2mm}\text{with}\hspace{2mm} \alpha\in \{x,y,z\}~.
	\end{aligned}
\end{equation}
The $u_{ij}$ are antisymmetric Hermitian link operators with eigenvalues $\pm 1$
\begin{equation}\label{eq:25}
	\begin{aligned}
		u_{ij} = -u_{ji}, \hspace{2mm} u^2_{ij} = 1, \hspace{2mm}, u^\dagger_{ij} = u_{ij}
	\end{aligned}
\end{equation}
The link operators commute with the Hamiltonian, $[H, u_{ij}] = 0$, so they are local symmetries. In this form, the Hamiltonian is representing a tight binding model of free Majorana fermions hopping along the lattice, with tunneling couplings that depend on the eigenvalues of link operators which can be thought of as a classical $\mathbb{Z}_2$ gauge field. One can assign a configuration $\{u\}$ to all link operators and diagonalize the resulting quadratic Hamiltonian in Majorana operators directly and obtain the spectrum and groundstate energy for $H\{u\}$. But which configuration $\{u\}$ corresponds to the global minimum of the energy? This question has been answered by Lieb (1994) \cite{Lieb_1994}. The groundstate lies in the sector with $B_p = 1 \hspace{2mm} \forall ~p$ and since the $B_p$ operators are the only gauge invariant objects, every gauge fixing choice $\{u\}$ which leads to this particular configuration for $B_p$ is equivalent. Therefore, our goal is to find a configuration $\{u\}$ which leads to $B_p = 1 \hspace{2mm} \forall ~p$. 

Next, we define the vortex operator $\hat{W}_p$ as
\begin{equation}\label{eq:26}
	\begin{aligned}
		\hat{W}_p = \prod_{(i,j)\in \partial P} u_{ij} \hspace{2mm}\text{where}\hspace{2mm}
		\left\{
		\begin{array}{l}
            i  \in \hspace{2mm} \text{even sublattice} \\
            j  \in \hspace{2mm} \text{odd sublattice}
        \end{array}
        \right.~.
	\end{aligned}
\end{equation}
The vortex operator can be simplified as below
\begin{equation}\label{eq:27}
	\begin{aligned}
		\hat{W}_p &= u^x_{12}u^y_{32}u^z_{34}u^x_{54}u^y_{56}u^z_{16} \\
		&= -(i)^6 c^x_1 c^x_2 c^y_2 c^y_3 c^z_3 c^z_4 c^x_4 c^x_5 c^y_5 c^y_6 c^z_6 c^z_1 \\
		&= \sigma^y_1 D_1 \sigma^z_2 D_2 \sigma^x_3 D_3 \sigma^y_4 D_4 \sigma^z_5 D_5 \sigma^x_6 D_6 ~,
	\end{aligned}
\end{equation}
where we used $i c^x_j c^y_j = -\sigma^z_j D_j$ and cyclic permutations. Considering the fact that $[\hat{W}_p, H] = [\hat{W}_p, D_i] = 0$, in the physical subspace where $D_j = 1$ we recover $\hat{W}_p = B_p$. Therefore, the uniform choice of $u_{ij}=1$ for every link will lead us to the groundstate in the extended (not yet projected) space with $B_p = 1$. In order to find the physical groundstate, we apply the global projector $\mathbf{D}$ to project out the unphysical states
\begin{equation}\label{eq:28}
	\begin{aligned}
		\ket{\psi_w}_{physical} = \mathbf{D} \ket{\psi_u}_{extended},  \hspace{2mm}\text{where}\hspace{2mm} \mathbf{D} = \prod^N_{i=1} P_i
		\end{aligned}
\end{equation}

\begin{figure}[!htb]
    \centering
    \includegraphics[width=0.45\textwidth]{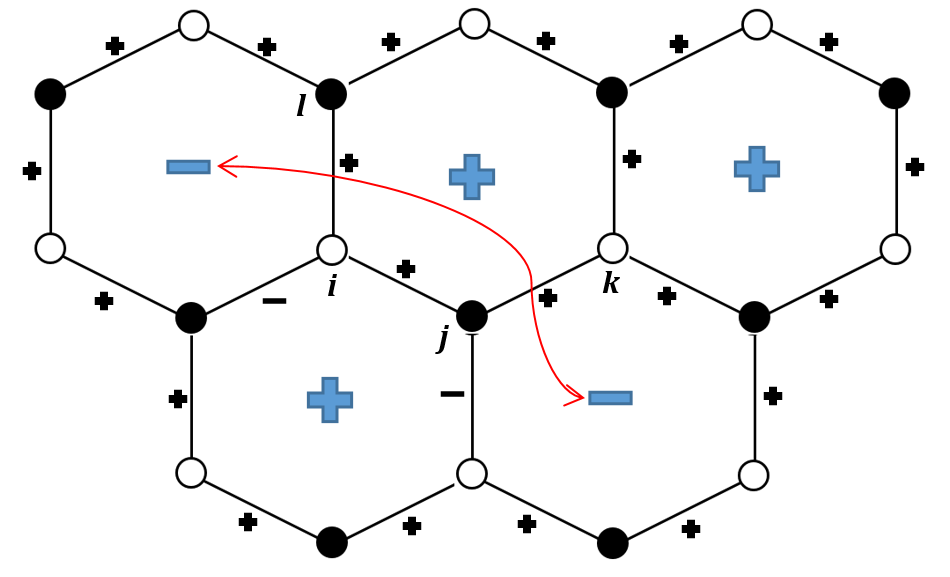}
    \caption{Two vortices connected by a string operator passing through several links. The white dots denote even lattice sites while the black dots are odd ones. By convention, the string operator is built up from operators acting on even sites.}
    \label{fig:string}
\end{figure}

Now we are at a point to explain in detail why the operators $\hat{O}_1$ and $\hat{O}_2$ can create vortices in the system. As mentioned earlier, if we flip the sign of the eigenvalue of the vortex operator at p, a vortex will be created in the system. Looking back at Eq. \eqref{eq:26}, we see that the vortex configuration $\{\hat{W}_p\}$ is created by fixing the gauge at every link $\{u_{ij}\}$. Therefore, by changing the link operators, we can move between vortex sectors. Starting from a vortex-free sector with all link operators being $+1$, applying $\hat{O}_1$ on site $a$ will leave the $\sigma^z_a$ untouched while flipping the $\sigma^x_a$ and $\sigma^y_a$. Consequently, this will change the sign of the eigenvalue of the $u_{ij}$ for x-link and y-link attached to site $a$, leaving the system with $B_{p_1} = -1$ and $B_{p_2} = -1$. We can also move the vortices around in the system, by consecutively applying these operators on the particular sites in a path for which we want to move the vortices along. As an example, the red path shown in the Fig. \ref{fig:string} corresponds to the operator
\begin{equation}
	\begin{aligned}
		\hat{O} = \exp{(-i\frac{\pi}{2}\sigma^z_i)} \exp{(-i\frac{\pi}{2}\sigma^x_j)}.
		\end{aligned}
\end{equation}

In general, we can define string operators to move the vortices anywhere in the system along the string, similar to the example above. To create such a string operator is straightforward now. If the string passes through a link, we need to apply an operator on one of the sites connected to that $\alpha-link$ which carries $\sigma^\alpha$ in the exponent. Before continuing, we need to make a convention. A sharp reader might point out that applying the operator $\hat{O}^{'} = \exp{(-i\frac{\pi}{2}\sigma^z_l)} \exp{(-i\frac{\pi}{2}\sigma^x_k)}$ instead of $\hat{O}$ creates the same pair of vortices in the system. For consistency reasons, we therefore use the convention that if a string is passing through a link, the corresponding string operator is built up by picking the even site on the link. Having said all of this,  a generic string operator is defined as

\begin{equation}\label{eq:29}
	\begin{aligned}
		\hat{S} = \prod_{\alpha-links} \exp{(-i\frac{\pi}{2}\hat{\sigma}^\alpha_\circ)}~.
		\end{aligned}
\end{equation}

This string operator will create two vortices at the ending points of the string which passes through the $\alpha-links$. The manipulation of link operators using these string operators is practically equivalent to changing the sign of the couplings $J_{ij}$ corresponding to the link $u_{ij}$ in the definition of the $A_{ij}$ (see Eq.  \eqref{eq:24}). Therefore, in the RBM language one can simply redefine the Hamiltonian with the desired signs of $J_{ij}$ for a particular vortex sector and then try to find the RBM representation for this new Hamiltonian through the approach explained in Section \ref{sec3}. The only practical disadvantage of this vortex realization approach is that it is computationally expensive as the RBM should be trained again from scratch to find the representation for the excited state. An ideal approach is one where, given the representation for the groundstate, we would be able to directly and without further optimization steps find the representation for the excited state. 

In principle, it is possible to directly manipulate the weight and bias parameters of the RBM,  $\Omega=(a_{k},b_{k'},W_{kk'})$ to realize the vortices and find the excited state representation. Looking at Eq.  \eqref{eq:19_1}, we notice that there is a symmetry between the eigenvalues of physical spins and the value of parameters of the RBM. In other words, flipping a particular spin in site $k$ is equivalent to flipping the sign of the RBM parameter sitting next to the $\sigma^{\alpha}_k$ in Eq.  \eqref{eq:19_1} and/or adding a phase to it. In general, for a string operator $\hat{S}_l^{\alpha} = \exp{(-i\frac{\pi}{2}\hat{\sigma}^{\alpha}_l)}$ we have

\begin{widetext}
\begin{equation}\label{eq:29_1}
	\begin{aligned}
		\hat{S}~|\Phi\rangle &= (-i\hat{\sigma}^{\alpha}_l)\sum_{\Xi}\Phi_{M}(\Xi;\Omega)|\Xi\rangle\\
		&= \sum_{\Xi} e^{[(-1)^{\delta(\alpha-z)-1} a_l]\sigma^{z}_l + \sum_{k\neq l} a_k \sigma_k^z} \times  
		\prod_{k'} \cosh &\left((-1)^{\delta(\alpha-z)-1} W_{lk'} \sigma_l^z + \sum_{k\neq l} W_{kk'} \sigma_k^z + b_{k'} \right) (-i\hat{\sigma}^{\alpha}_l)~|\Xi\rangle~.
	\end{aligned}
\end{equation}
\end{widetext}

Next, one needs to evaluate how $(-i \hat{\sigma}^{\alpha}_a)$ acts on $|\Xi\rangle$. Here we subdivide the three cases $\alpha =z$, $x$ and $y$. If $\alpha =z$, spin up states will get a phase $-i$ and spin down states a phase $i$. The overall effect of this can be absorbed in a shift of $a_l$ by $-i \frac{\pi}{2}$. If $\alpha=x$, then spins up and down are just exchanged and the $-i$ is an overall phase which can be ignored. If $\alpha = y$, we get a combination of the previous two cases. In equation form, the above explanations take the following form
\begin{equation}
\label{eq:29_2}
	\begin{aligned}
		\hat{S}^z_l & : \left(W_{lk'},a_l\right) \mapsto \left(W_{lk'},a_l- i \frac{\pi}{2}\right) \\	
    	\hat{S}^x_l & : \left(W_{lk'},a_l\right) \mapsto \left(-W_{lk'},-a_l\right) \\
    	\hat{S}^y_l & : \left(W_{lk'},a_l\right) \mapsto \left(-W_{lk'},-a_l-i \frac{\pi}{2}\right)~.
	\end{aligned}
\end{equation}

\section{Results and Discussion}\label{sec5}

\subsection{Groundstate}

In this section, we present the results for the groundstate energy estimate of different system sizes for the proposed mapping discussed in Section \ref{sec3}. These results are also compared to the exact analytical expression given in Eq.  \eqref{eq:17}. The absolute value of the energies is plotted in Fig. \ref{tab:r_space} in Appendix \ref{sec8}. All results for this and the following sections are for the phase $J_x = J_y = J_z = 1$.

We used both \texttt{NetKet} and PyTorch to build our RBM. While \texttt{NetKet} provides build-in functions to evaluate physical observables, we build our own version of these functions in the PyTorch environment. When training with \texttt{NetKet}, the configuration space of the network is sampled for hundreds of  times, depending on the lattice size, using the Metropolis algorithm. While when training with PyTorch, the configuration space of the network is sampled from all of the Hilbert space. This is easier to implement and enables us to focus on the explicit RBM representation. The loss function of the network has been minimized using Stochastic Gradient Descent and the learning rates are set to decay using the Adaptive Moment (ADAM) estimation algorithm. The groundstate energies are obtained by fine-tuning the optimisation algorithm parameters.

The numerical results obtained are shown in Fig. \ref{tab:r_space} in Appendix \ref{sec8}. The different methods for computing the groundstate energy listed in the Table  \ref{tab:0} are as follows:
\begin{enumerate}
	\item[a)] Analytic formula Eq.  \eqref{eq:17} using
	\begin{itemize}
		\item[i.] anti-periodic boundary conditions
		\item[ii.] periodic boundary conditions
	\end{itemize}
	\item[b)] The method used by Kitaev and Pachos \cite{Kitaev_2003,pachos_2012}
	\item[c)] Direct diagonalization for small system sizes, using \texttt{NetKet}.
	\item[d)] RBM results for an architecture with $\alpha = \frac{\textrm{num}_{h}}{\textrm{num}_{v}} = 2$ and 1k samples and 10k iteration steps, using \texttt{NetKet}.
	\item[e)] RBM results built using PyTorch with sampling all of the Hilbert space and using projection onto translation invariant states as described below.
\end{enumerate}

\begin{table*}[!htb]
	\centering
    {\rowcolors{2}{green!80!yellow!50}{green!60!yellow!30}
    \begin{tabular}[c]{c|c c c c c c c c c c}
    lattice size&$2\times2$    &$2\times3$    &$2\times4$     &$3\times3$     &$3\times4$     &$4\times4$     &$5\times5$     &$6\times6$     &$7\times7$ &$3\times3$ with $K=0.2$\\
    \hline 
    a.i         &-6.4721\footnotemark[1]
    &-9.8003
    &-12.5851\footnotemark[1]
    &-14.2915&-18.9869&-25.1282&-39.3892&-56.7529&-77.1249&\\
    a.ii        &-6     &-9.2915&-12.4721&-14.2915&-19.0918&-25.4164&-39.3892&-56.2668&-77.1249&\\
    b           &-6     &-9.2915&-12.4721&-14.2915&-19.0918&-25.4164&-39.3892&-56.2668&-77.1249&\\
    c           &-6.9282&-9.8003&-12.9443&-14.2915&-19.0918&-&-&-&-&-17.5260\\
    d           &-6.9282&-9.8003&-12.573 &-13.7374&-18.0466&-24.0783&-37.305 &-52.920 &-71.793&\\ 
    e           &-6.9282&-9.8003&-12.9293&-14.2787& & & & & &-17.393
    \end{tabular}}
    \footnotetext[1]{Notice that the energies obtained via the analytical formula do not match the exact diagonalization results for $2\times2$ and $2\times4$ lattices. This is not a problem because the energy obtained via the analytical formula (method a.i) is still in the spectrum, which means this is a case where Lieb’s theorem fails - the vortex free sector no longer contains the ground state. This is an edge case and will not appear when $m,n>2$.}
	\caption{\label{tab:0} Groundstate energy for different system sizes obtained via different methods explained in the text.}
	
\end{table*}

Based on the results obtained for the energy of the groundstate of the system, training of the network is tested and optimized by changing the learning rate for different training phases and iteration numbers. The most effective initial learning rate is found to be $\textrm{lr}_{init}\approx0.01$. This value is small enough to avoid over-fitting and large enough to train the network effectively. This learning rate is rather large compared to standard machine learning implementations, because the number of parameters in our RBM is relatively small compared to hundreds of millions of parameters used in typical deep learning applications. 

A central ingredient of our approach in PyTorch is to use  \textit{projection} onto translation invariant states in order to restrict RBM parameters to the physical parameter space. This dramatically improves the results we can achieve. The theoretical reason for this projection is that ground states of our Hamiltonian reside in such translation invariant subspaces of the Hilbert space as explained in Section \ref{sec:cft}. Here we provide the technical details of the implementation. Suppose $|\Phi\rangle$ is the RBM state mentioned in Eq.  \eqref{eq:19_0}. Then the state used in calculating the energy is as follows
\begin{equation}
    |\Phi'\rangle=\sum_{m,n}\hat{T}_{m\textbf{a}_1+n\textbf{a}_2}|\Phi\rangle~,
\end{equation}
where $\hat{T}_{m\textbf{a}_1+n\textbf{a}_2}$ are operators translating $m$ steps along the $\textbf{a}_1$ direction and $n$ steps along the $\textbf{a}_2$ direction, with $\textbf{a}_1$, $\textbf{a}_2$ defined in Eq.  \eqref{eq:0}. $\hat{T}_{m\textbf{a}_1+n\textbf{a}_2}$ can be realized either as a permutation matrix acting on quantum states or a permutation of parameters in RBM.

The hyperbolic trigonometric functions appearing in the weights of the RBM are numerically unstable. They sometimes lead to divergence and we have to use an extremely small learning rate to avoid the divergence. To cure this instability, we use the following \textit{pre-train} algorithm. We first perform the substitution
\begin{equation}
    \cosh x\mapsto 1+\frac{x^2}{2},\ \sinh x\mapsto x~,
\end{equation}
as the polynomials $x$ and $x^2$ are more stable under numerical changes than the exponential $e^x$ which appears in the hyperbolic trigonometric functions. Then this modified model is trained for several thousand epochs. After we get close enough to the minimum, we change back to the standard RBM representation using $\cosh$ and $\sinh$ to refine our result. We find that this two-step training shortens the time to reach a given accuracy.

Remember that RBM parameters are complex numbers. However, the meaning of the real part and the imaginary part of parameters are different. The real part contributes by $\cosh/\sinh$ while imaginary parts show  up as arguments of $\cos/\sin$, i.e. they are phases. So they should be initialized with different standard deviations in the sampling distribution and trained with different learning rates. In other words they are ``renormalized'' differently along the flow. 

On the other hand, when working with \texttt{NetKet}, we find that the ground state energy cannot be reached beyond a certain accuracy. Thus the results obtained with our PyTorch algorithm described above are always far more accurate than the best results \texttt{NetKet} can achieve. We believe this is due to the following identity 
\begin{equation} \label{eq:secmult}
\Pi_i x_i=e^{\sum_i \log x_i}~,  
\end{equation}
used by \texttt{NetKet}, in order to accelerate sequential multiplication. This is theoretically correct but numerically unstable as in an RBM representation there are always extremely large or small values of $\cosh/\sinh$ (this is the means by which an RBM is capable of approximating the correct physical state!) leading to truncation errors when using the functions $\log$ and $\exp$. So it is not surprising that \texttt{NetKet} cannot reach a high enough accuracy and gets stuck at a too high energy. We find that when using the identity (\ref{eq:secmult}) in our PyTorch implementation, our ground state energy becomes similarly inaccurate as the one obtained by \texttt{NetKet}. 

In order to check the effect of the relative number of hidden and visible nodes on the outcome of the network, different values for $\alpha$ were tested, keeping other parameters fixed. The $\alpha$ parameter was changed just by changing the number of hidden nodes for a fixed system size (and hence fixed visible layer size). The results (for a $5\times 5$ lattice) are summarized in the Table \ref{tab:alpha} in Appendix \ref{sec8}.

\vspace{4mm}
\textbf{- The performance of FFNN vs. no-visible-bias RBM}

In Table  \ref{tab:ffnn} in Appendix \ref{sec8}, we give a brief summary of the comparison results between an FFNN with one layer and a no-visible-bias RBM (referred to as novb-RBM in the table). The two architectures are similar and for the same computational resources (such as CPU and memory) and system size these results have been obtained. The conclusions we reach are:
\begin{enumerate}
	\item The FFNN trains faster than the RBM with relatively fewer free parameters and performs better than the RBM with no bias.
	\item But the FFNN's result deviates by a greater margin from the analytic result.
\end{enumerate}

\vspace{4mm}
\textbf{- RBM's parameters: an example}

To get a better feeling for how the RBM approximates quantum states so well, let us have a closer look at an RBM for a small system size, as shown in Eq.  \eqref{eq:rbm_paras}. These are the parameters for an RBM with 5 hidden nodes for a $2\times2$ system which has 8 spinors. So the parameter matrix $W_{kk'}$ is $5\times8$ in size. The indexing rule for the spinors is shown in Fig. \ref{qst_vor}. The biases $a_k$ and $b_{k'}$ in RBM are set to $0$ manually. For illustration purposes, this RBM is trained without the \textit{projection method}. After training, the energy of the resulting parameter set is $-6.928201$ while the true ground state energy is $-6.92820323$. Its wave-function has $99.999982\%$ overlap with the groundstate.

\begin{widetext}
\begin{equation}\label{eq:rbm_paras}
\begin{aligned}
\Re(W)&=\begin{bmatrix}
     0.3399&  0.3517&  0.3000& -0.2862&  0.2862&  0.3000& -0.3516&  0.3398\\
     0.0000&  0.0000&  0.0000&  0.0000&  0.0000&  0.0000&  0.0000&  0.0000\\
    -0.3687&  0.3906& -0.6170& -0.5241&  0.5241& -0.6170& -0.3906& -0.3687\\
     0.3236& -0.3456&  0.5754&  0.4805&  0.4805& -0.5754& -0.3456& -0.3236\\
     0.0000&  0.0000&  0.0000&  0.0000&  0.0000&  0.0000&  0.0000&  0.0000
\end{bmatrix}\\
\Im(W)&=\begin{bmatrix}
     0.7658&  0.7894& -0.7439& -0.7592& -0.0264&  0.0412& -0.0037& -0.0199\\
     0.0000&  0.0000&  0.0000&  0.0000& -0.7854& -0.7854&  0.7854&  0.7854\\
     0.2733& -0.4073&  0.4872&  0.1436&  0.6418& -0.2982& -0.3781& -0.5121\\
     0.2612&  1.0566& -0.7134& -0.0218& -0.8072& -0.0720&  0.2712&  0.5242\\
     0.7854&  0.7854& -0.7854& -0.7854& -0.7854& -0.7854&  0.7854&  0.7854
\end{bmatrix}
\end{aligned}
\end{equation}
\end{widetext}

To begin with, notice that, for example, for a state with all spins up, the RBM amplitude is given by
\begin{equation}
    \Phi_M(|\uparrow\uparrow\uparrow\uparrow\uparrow\uparrow\uparrow\uparrow\rangle) = \prod_{k'=1}^5 \cosh\left(\sum_{k=1}^8 W_{kk'}\right),
\end{equation}
where we are using the convention that spin up corresponds to the value $\sigma^z=1$ and spin down to $\sigma^z = -1$.
If we focus on the factor corresponding to $k'=5$ and use the identification $\pi/4\approx0.7854$, we see from Eq.  \eqref{eq:rbm_paras} that its amplitude is given by
\begin{equation}
    \cos\left(\frac{\pi}{4} + \frac{\pi}{4} - \frac{\pi}{4} - \frac{\pi}{4} -\frac{\pi}{4} -\frac{\pi}{4} +\frac{\pi}{4} + \frac{\pi}{4} \right) = \cos(0) = 1,
\end{equation}
where we have used that
\begin{equation}
    \cosh(x+iy)=\cosh(x)\cos(y)+i\sinh(x)\sin(y).
\end{equation}
On the other hand, if we flip one of the spins, the corresponding amplitude becomes
\begin{eqnarray}
    ~ & ~ & \Phi_M(|\uparrow\uparrow\uparrow\uparrow\downarrow\uparrow\uparrow\uparrow\rangle) \nonumber \\
    ~ & = & \cos\left(\frac{\pi}{4} + \frac{\pi}{4} - \frac{\pi}{4} - \frac{\pi}{4} +\frac{\pi}{4} -\frac{\pi}{4} +\frac{\pi}{4} + \frac{\pi}{4} \right) \nonumber \\
    ~ & = & \cos\left(\frac{\pi}{2}\right) = 0.
\end{eqnarray}
That means the second and fifth line of $\Im(W)$ impose strong restrictions on spinor configurations and in this particular case imply that up and down spins are paired in even numbers. This restriction also happens in the first, third and fourth line of $\Im(W)$, although this is not apparent from first sight. This can be seen from observing that
\begin{equation}
\begin{aligned}
\Im(W)[0,0]-\Im(W)[0,7]&=0.7857\approx\pi/4~, \\
\Im(W)[0,3]-\Im(W)[0,4]&=-0.7856\approx-\pi/4.
\end{aligned}
\end{equation}

\begin{figure}[!ht]
	\centering
	\includegraphics[width=0.25\textwidth]{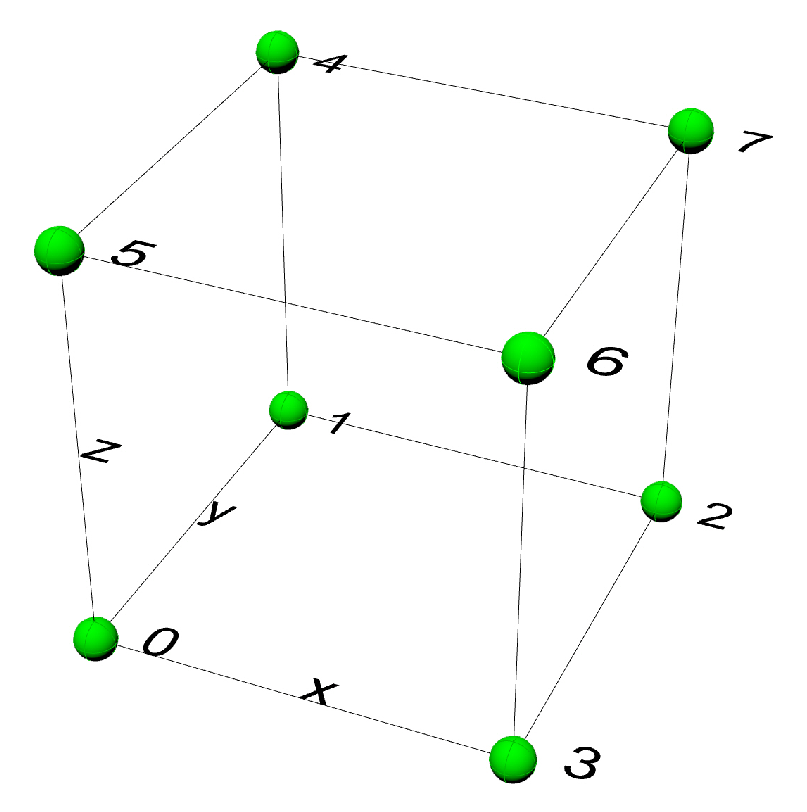}
	\caption{$2\times2$ honeycomb lattice can be drawn as a cubic. The indexing rule for the spinors is shown in Fig. \ref{qst_vor}.}
	\label{fig:cubic2x2}
\end{figure}

Furthermore, there are many numerical coincidences for parameters between spinor pairs $(0,7)~(1,6)~(2,5)$ and $(3,4)$. This reflects the “parity” symmetry of the $2\times 2$ lattice. More specifically, the $2\times2$ lattice can be drawn as a cubic as shown in Fig. \ref{fig:cubic2x2}. Then it is apparent that there are strong correlations between the pairs $(0,7)~(1,6)~(2,5)$ and $(3,4)$.

This small example shows that an RBM is able to find restrictions and symmetries of our quantum system efficiently. We can imagine that this will also happen for larger systems, although it is hard to observe such symmetries explicitly since the number of parameters becomes too large.

\subsection{Creating Vortex pairs and measuring their energy}

As described in Section \ref{sec4}, there are three theoretically equivalent methods to create vortices:
\begin{enumerate}
    \item[a)] modifying parameters of RBM;
    \item[b)] creating an auxiliary Hamiltonian and training it again;
    \item[c)] transforming into auxiliary fermion representation and fixing $u_{ij}$s \cite{Kitaev_2003,pachos_2012}.
\end{enumerate}
We applied all three methods to a 3x3 lattice and computed the corresponding energy levels for excited states. The results we get are shown in Table \ref{tab:3x3}.

\begin{table}[!htb]
	\centering
	{\rowcolors{3}{green!80!yellow!50}{green!60!yellow!30}
    \begin{tabular}{c|c c c }
    \multirow{2}{*}{Spins Flipped}&Energy&Energy&Energy\\
    &(Method A)&(Method B)&(Method C)\\
    \hline 
    $\sigma_8^y$&-11.7335(5) &-12.1070(51)&-13.9146\\
    $\sigma_9^y$&-11.7474(5) &-11.8873(57)&-13.9146\\
    $\sigma_8^x\sigma_8^z$&unrealizable&-12.7447(37)&-13.9146\\
    \end{tabular}}
	\caption{Energy of excited states with a pair of vortices in a $3 \times 3$ lattice system, measured through different methods as explained in the text.}
	\label{tab:3x3}
\end{table}

We find that method (a) is more stable than method (b), which might be due to the random initialization of RBM parameters in method (b) for the training process. Both methods, (a) and (b), deviate by a margin from the analytic values, namely method (c). This can be explained by the fact that the state generated by the spin flip operation is not the groundstate of the vortex sector but rather an excited state carrying quasi-particles. We will comment on this further in the next sub-sections about the expectation values of plaquette operators.

\subsection{Perturbation with non-trivial magnetic field}
\label{sec:nonzeroK}

As originally discussed by Kitaev \cite{Kitaev_2006}, one can add the following perturbation to the Hamiltonian \eqref{eq:1} 
\begin{equation}\label{eq:Bfield}
    \Delta H = - K \sum_{(i,j,k)} \sigma_i^x \sigma_j^y \sigma_k^z,
\end{equation}
which corresponds to an effective magnetic field with coupling $K$, and the sum is over all vertices with links $(i,j,k)$. This term leads to an energy gap in the spectrum for asymptotically large lattices and renders the vortex sectors of the model topological. In this phase, the vortices of the honeycomb lattice bind localized Majorana modes and behave as Ising anyons. For the $3\times 3$ lattice, we find that the gap in the spectrum separating the groundstates from excited states becomes larger for non-zero $K$. For $K=0.2$, exact diagonalization gives a GS energy of $-17.5260$. We find that using our RBM approach, with the same number of parameters and training procedures discussed above, we get an energy value of $-17.393$ (0.75\% deviation) which is quite close to the analytical value, as shown in Table \ref{tab:0}. This shows that the RBM can be an efficient and accurate representation for the non-trivial topological phases of the Kitaev honeycomb model. As in this work we are mainly interested in the groundstate, which already displays topological order for the $3\times 3$ lattice, we leave further work on the phase with non-trivial $K$ for the future.

\subsection{Quantum State Tomography for Groundstate}\label{sec5.3}

\begin{figure}[!htb]
	\centering
	\includegraphics[width=0.48\textwidth]{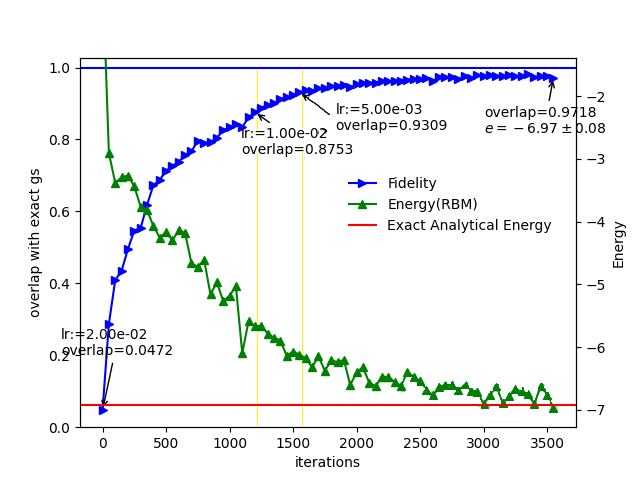}
	\caption{\label{fig:qst} Energy and overlap results for QST of $2 \times 2$ lattice system for different training phases with specified learning rates in the figure.} 
\end{figure}

In this section we apply the technique discussed in Section \ref{sec3.2} to reconstruct the wave function from measurements . In order to generate measurement data set, we sample from an exact groundstate by random measurements in different basis. To this end, we focus on the $2 \times 2$ lattice in order to deal with the computational resource requests and also being able to find the exact groundstate wave function through exact diagonalization. The last step is to train the RBM using the data set obtained. The result is shown in Fig. \ref{fig:qst}.

There are many intricate parameters which one can adjust during training, such as learning rate, batch size, number of Metropolis samples, and number of single measurements. Only with a suitable choice of parameters and large enough training data set, can one achieve a satisfactory result. For the example depicted in Fig.  \ref{fig:qst}, in order to build the training data set, the exact wave function was sampled 2000 times for 256 randomly chosen basis $x, y, z$ and the hidden node density $\alpha$ was set equal to $2$. With this combinations of parameters, QST leads to an overlap of 97 percent between the exact wave function and the learned one. 

\vspace{4mm}
\textbf{- Plaquette Operator expectation values and spin flipping}

\begin{figure}[!htb]
	\centering
	\includegraphics[width=0.25\textwidth]{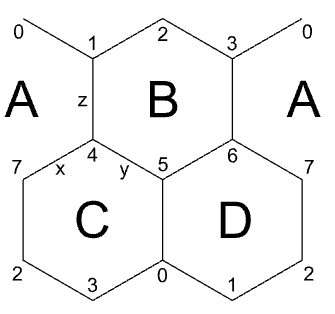}
	\caption{\label{qst_vor} Labeling convention for plaquette operator measurements.} 
\end{figure}

Starting from the state obtained from tomography as discussed above, one can then realize vortices in the system by modifying parameters of the RBM as described in Section \ref{sec4}. In order to verify that modification of RBM parameters takes us to the right excited state, we measure the relevant plaquette operators in the system. The convention for labeling plaquette operators and sites is as shown in Fig.  \ref{qst_vor}. 

For the trained RBM state after performing QST for groundstate of $2 \times 2$ lattice, the expectation values of its four plaquette operators as well as the measured energies are shown in Table \ref{tab:pl0}. Moreover, flipping the $y$- and $z$-components of the spin at lattice site number $5$ (or equivalently, applying $\hat{S}_5^{x}$) will produce a pair of vortices at $B$ and $D$ plaquettes. Therefore, the expectation value of the $\hat{W}_p$ operator in these plaquettes also flips sign. Similarly, applying $\hat{S}_7^{y}$, will kill the vortex in plaquette $D$ and produce one in plaquette $A$. The changes in the sign of expectation values for these plaquettes is shown in the Table  \ref{tab:pl0}. Finally, by applying $\hat{S}_3^{z}$ we return back to the no-vortex sector. These results, prove the effectiveness of the framework developed in Section \ref{sec4} for realization of vortices in RBM states, in both QST and usual learning approaches. 

Note that the energy after each flip changes by an amount $\Delta E \sim 2$. This can be explained by the fact that each vortex sector has its own groundstate but also excited states with higher energies created by fermionic quasi-particle excitations. For example, the no-vortex sector has a spectrum given by the Hamiltonian in Eq.  \eqref{eq:17} and while the groundstate is the lowest lying state in this sector, there will be excited quasi-particle states carrying momentum $\mathbf{k}$ which are created by the $\gamma^{\dagger}_{\mathbf{k}}$ operators. Similarly, vortex sectors with a non-trivial vortex number can be in an excited state by applying quasi-particle creation operators. A rough estimate shows that the difference $\Delta E \sim 2$ is of the order of several such a quasi-particle excitations and thus the vortices we are creating carry such excitations. This also explains why when returning to the no-vortex sector we do not arrive at the original energy.
\begin{table}[!htb]
    \centering
    {\rowcolors{3}{green!80!yellow!50}{green!60!yellow!30}
    \begin{tabular}{c|ccccc}
         Spin Flip Operator&\multirow{2}{*}{$W_A$}&\multirow{2}{*}{$W_B$}&\multirow{2}{*}{$W_C$}&\multirow{2}{*}{$W_D$}&\multirow{2}{*}{Energy}  \\
         (acting successively)& \\
         \hline
         Groundstate&0.9489&0.9517&0.9525&0.9553&-6.078\\
         $\hat{S}_5^{x}$&0.9489&-0.9517&0.9525&-0.9553&-4.011\\
         $\hat{S}_7^{y}$&-0.9489&-0.9517&0.9525&0.9553&-1.804\\
         $\hat{S}_3^{z}$&0.9489&0.9517&0.9525&0.9553&0.061\\
    \end{tabular}}
    \caption{Expectation values of plaquette operators after spin flips.}
    \label{tab:pl0}
\end{table}

\section{Relation to Conformal Blocks and Moore-Read state}
\label{sec:cft}

As originally described in \cite{Kitaev_2006}, the non-Abelian sector of the Kitaev Honeycomb model which we are addressing in the current paper is in the same universality class as the Ising conformal field theory. That is the topological excitations of the model correspond to the primary operators of the chiral sector of the Ising CFT, namely the vacuum corresponds to the identity operator $\mathbb{1}$, the vortex corresponds to the $\sigma$-field and the quasi-particle corresponds to the $\psi$-field. This relation is part of a broader framework connecting topological phases of $2+1$-dimensional quantum field theories to chiral sectors of WZW conformal field theories that can be traced back to the work of Witten on Chern-Simons theory and WZW models \cite{Witten:1988hf,Witten:1991mm}, see \cite{Nayak_2008} for a review. Within this correspondence wave-functions of the $2+1$-dimensional topological theory are mapped to conformal blocks of the WZW model and vice versa. One very well-known application of this paradigm is the description of fractional Quantum Hall states with filling fraction
\begin{equation}
    \nu = \frac{p}{k+pn}, 
\end{equation}
using conformal blocks in an $SU(p)_k$ WZW model \cite{Blok:1991zq}, where $p$ and $k$ are positive integers, and $n$ is an arbitrary positive integer. For $p=1$, these are simply the Laughlin states. 
The case of Ising anyons corresponds to the choice $p=k=2$ and leads to the famous Moore-Read state \cite{Moore:1991ks}. 

Let us now describe this map in some degree of detail which is suitable for our purposes. The ground state wave-function of the topological theory on the 2-sphere is described by the identification \cite{Dorey:2016mxm}
\begin{widetext}
\begin{equation}
    \Psi_0(z_1,\ldots,z_N) = \prod_{i<j}^N (z_i-z_j)^{n+k/p} \langle \mathcal{O}_R(z_1) \ldots \mathcal{O}_R(z_N)\rangle e^{- \sum_i |z_i|^2/4l_B^2},
\end{equation}
\end{widetext}
where in order to obtain a unique ground state, the conformal block $\langle \mathcal{O}_R(z_1) \ldots \mathcal{O}_R(z_N)\rangle$ is computed in the representation $R = \textrm{Sym}_k$, namely the $k$th symmetric representation of $SU(p)$, which has the property
\begin{equation}
    \underbrace{\textrm{Sym}_k \times \textrm{Sym}_k \times \ldots \times \textrm{Sym}_k}_{p} = \textbf{1}.
\end{equation}
In the case of quantum hall states, $l_B = \sqrt{\hbar/eB}$ and the $z_i$ are positions of ``electrons''. Specifying to the case $p=k=2$, we obtain the \textit{Pfaffian} state \cite{Dorey:2016mxm}
\begin{equation} \label{eq:spherepf}
    \Psi^{\textrm{Pf}}_0(\{z_i\}) = \textrm{Pf}\left(\frac{|ij\rangle_1}{z_i-z_j}\right) \prod_{i<j} (z_i-z_j)^{n+1} e^{-\frac{1}{4l_B^2}\sum_j |z_j|^2}~,
\end{equation}
where we have taken $\mathcal{O}_R$ to be the spin $1$ field $\psi$ of the Ising model and the Pfaffian of a $2m$-by-$2m$ antisymmetric matrix is defined as
\begin{equation}
    \textrm{Pf}(A) = \frac{1}{2^m m!} \sum_{P\in S_{2m}} \textrm{sgn}(P) \prod_{i=1}^m A_{P(2i-1),P(2i)}~.
\end{equation}
The state $|ij\rangle_1$ appearing in the Pfaffian is a spin singlet formed from two spin $1$ states
\begin{equation}
     |ij\rangle_1 = |1_i\rangle |-1_j\rangle + |-1_i\rangle |1_j\rangle - 2 |0_i\rangle |0_j\rangle.
\end{equation}
We will see the significance of this choice shortly. The non-Abelian sector of the Kitaev model corresponds to the value $n=0$ as quasi-particles carry the same electric charge as an electron 
\begin{equation}
     e^* = \nu e = e \quad \Rightarrow \quad \nu = 1 \quad \Rightarrow \quad n = 0.
\end{equation}
With these choices equation Eq.  \eqref{eq:spherepf} is a spin singlet version of the Moore Read state \cite{Moore:1991ks}. Next, we would like to interpret Eq.  \eqref{eq:spherepf} as a ground state wave-function of the Kitaev model. That such a correspondence is possible for spin lattice models has been previously shown in the case of Laughlin spin-liquid states on lattices \cite{PhysRevLett.108.257206,Glasser_2018}. Passing over to the torus with periodic boundary conditions, which is the situation examined in the present paper, the Pfaffian state lifts by the replacements 
\begin{equation}
    \frac{1}{z_i-z_j} \mapsto \frac{\vartheta[\alpha](z_i-z_j)}{\vartheta_1(z_1-z_j)}, \quad z_i = (x_i + i y_i)/L_x,
\end{equation}
to the following expression involving theta functions \cite{Chung_2007}, 
\begin{widetext}
\begin{equation} \label{eq:ansatz}
    \Psi^{\textrm{Pf}}_{0,\alpha}(z_1,\ldots,z_N) = e^{-\sum_i |z_i|^2/4l_B^2} F_{\textrm{cm}}
^{\alpha}(\sum_i z_i) \textrm{Pf}\left(\frac{\vartheta[\alpha](z_i-z_j)}{\vartheta_1(z_i-z_j)}|ij\rangle_1\right)\prod_{i<j} \vartheta_1(z_i-z_j),
\end{equation}
\end{widetext}
where the modular parameter of the theta-functions is chosen to be the one of the lattice\footnote{After publishing this work as preprint, we were notified by the authors of \cite{PhysRevB.103.075130} who had obtained a similar ansatz for rectangular lattices with periodic boundary conditions.}, namely in our case $\tau = e^{\pi i/3}$. The center of mass contribution has been modified in order to accommodate for the fact that $\nu=1$ and, using  $N_s = L_x L_y/2\pi l_B^2$, takes the form
\begin{equation}
    F_{\textrm{cm}}^{\alpha}(z) = \vartheta\left[(N_s-2)/2+(1-2\epsilon)/2\atop -(N_s-2)/2-(1-2\epsilon')/2\right](z,\tau),
\end{equation}
and we refer the reader to Appendix \ref{sec:theta} for conventions and definitions of the theta functions. Note that on the torus, the ground state is triply degenerate with the different states corresponding to the values $\alpha=(1/2,0)$, $(0,0)$, and $(0,1/2)$. The transition between these ground states can be understood as creation of vortex/anti-vortex pairs which are then successively transported along the cycles of the torus and then annihilated, as beautifully described in \cite{Oshikawa_2007}. Moreover, for $B=0$ we have $l_B \rightarrow \infty$ and $N_s=0$, resulting in periodic boundary conditions for the wave-function. 

Now let us make the connection between what we have said above about the Moore-Read state and the RBM ground state wave-functions constructed in this paper. First of all, note that the RBM wave-function is in the spin basis of the electrons at each lattice site. Recall that, as discussed in Section \ref{sec4}, the Hilbert space of our spin $\frac{1}{2}$ electrons can be described by two spinless complex fermions $a_1$ and $a_2$. In particular, spin up will correspond to both modes unoccupied, while spin down arises from both modes occupied:
\begin{equation} \label{eq:spinrep}
    |\uparrow\rangle = |00\rangle, \quad |\downarrow\rangle = |11\rangle,
\end{equation}
with $a_1 |00\rangle = a_2 |00\rangle = 0$ and $|11\rangle = a_1^{\dagger}a_2^{\dagger} |00\rangle$. However, we remark here that $|\uparrow\rangle$ and $|\downarrow\rangle$ are not the standard spin up and down basis vectors with respect to $\sigma^z$, but rather rotated basis vectors which respect the symmetries of the Honeycomb Hamiltonian which for $J_x=J_y=J_z=1$ is invariant under permutation of $x$, $y$ and $z$. In particular, we make the choice of basis satisfying
\begin{eqnarray}
    \langle \uparrow | \sigma^{\alpha} | \uparrow \rangle = \frac{1}{\sqrt{3}}, \quad \langle \downarrow | \sigma^{\alpha} | \downarrow \rangle = -\frac{1}{\sqrt{3}}~,
\end{eqnarray}
where $\alpha = z,x,y$. A concrete solution in the $\sigma^z$-basis satisfying the above constrains is given by
\begin{equation}
    |\uparrow\rangle = \left(\sqrt{\frac{1}{2}+\frac{1}{2\sqrt{3}}}\atop \sqrt{\frac{1}{2}-\frac{1}{2\sqrt{3}}}~e^{i \pi/4}\right),~  
    |\downarrow\rangle = \left(\sqrt{\frac{1}{2}-\frac{1}{2\sqrt{3}}}\atop \sqrt{\frac{1}{2}+\frac{1}{2\sqrt{3}}}~e^{i 5\pi/4}\right).
\end{equation}
The representation \eqref{eq:spinrep} is faithful if we restrict to the subspace of fermionic states $|\Psi\rangle$ that satisfy
\begin{equation}
    D_i |\Psi\rangle = |\Psi\rangle, \quad D_i = (1-2a_{1,i}^{\dagger}a_{1,i})(1-2a_{2,i}^{\dagger}a_{2,i}).
\end{equation}
This projects out the unphysical states $|10\rangle$ and $|01\rangle$. As pointed out in \cite{Dorey:2016mxm}, the unprojected system can also describe the degrees of freedom of a spin $1$ particle
\begin{equation}
    |00\rangle = |1\rangle, \quad |11\rangle = |-1\rangle, \quad |10\rangle = |01\rangle = |0\rangle. 
\end{equation}
Thinking reversely, we propose to identify our RBM state $|\Phi\rangle$ with the Pfaffian state by projecting onto physical states \footnote{In general, the RBM state obtained after training will be a linear combination of the $3$ groundstates.}
\begin{equation} \label{eq:MRphys}
    |\Phi\rangle \sim \widehat{P}_{\textrm{physical}} \Psi^{\textrm{Pf}}_{0,\alpha},
\end{equation}
where the positions $z_i$ of spin $1$ operators $\psi$ are identified with complex lattice point coordinates of lattice site $i$ on the torus. Put another way, an electron at site $i$ of the Honeycomb lattice is represented in the conformal block by $\psi(z_i)$. Note that the number of lattice sites is twice the number of plaquettes and therefore always even such that the Pfaffian is well-defined. Moreover, note that the projection operator $\widehat{P}_{\textrm{physical}}$ projects our spin $1$ singlet onto the following spin $\frac{1}{2}$ tensor product
\begin{equation} \label{eq:spinprojection}
    \bigotimes_k \left(\frac{1+D_k}{2}\right) |ij\rangle_1 = |\uparrow_i\rangle |\downarrow_j\rangle + |\downarrow_i\rangle |\uparrow_j\rangle.
\end{equation}
The relation \eqref{eq:MRphys} should not be thought of as an exact equality but rather as an approximation. The reason is that due to the complexity of the Pfaffian state it is not possible for us, yet, to show that the projected state respects all the symmetries of the Honeycomb Hamiltonian. In fact, our Ansatz \eqref{eq:ansatz} is highly symmetric and is invariant under the $120$ degree rotational symmetry of the Honeycomb lattice as well as the symmetry operator which permutes the Pauli matrices, as described in Appendix \ref{sec:PFapp}. Together, these symmetries leave the Hamiltonian invariant and should thus also leave the groundstate sector invariant, which in Appendix \ref{sec:PFapp} is used to conjecture a more precise version of the correspondence \eqref{eq:MRphys}. Moreover, we note that our Ansatz satisfies three key criteria which the ground state wave-function has to respect. Namely, it is invariant under translations by torus lattice generators, it is triply degenerate, and it is lying in a $2^m$-dimensional sub-space of the $2^{2m}$-dimensional Hilbert space which also happens to be the dimension of the vortex-free sector. 

\section{Conclusion}\label{sec6}
\label{sec:conclusion}

In this paper, we have constructed RBM realizations of several sectors of the Kitaev Honeycomb model. In particular, we have focused on the gap-less non-Abelian phase corresponding to the parameter choice $J_x = J_y = J_z$ and used machine learning algorithms to train restricted Boltzmann machines to find groundstates and excited states of such systems. We used two different approaches to find these states, namely \texttt{NetKet} and PyTorch. We find that using PyTorch and our own software package, we can achieve a very accurate result for the groundstate energy for small lattice sizes. On the other, the RBM trained with \texttt{NetKet} does not give an accurate enough result for the energy but the sampling algorithm employed is very efficient and allows \texttt{NetKet} to also proper larger lattices. An ideal approach would be to combine \texttt{NetKet}'s sampling algorithm with our own groundstate search algorithm. We leave this task for the future. Since our RBM wave-functions have $99.8\%$ overlap with the exact groundstate (this result is for the $3\times3$ lattice), we find that we can transition to vortex sectors by manipulating RBM parameters as described in Section \ref{sec4}. This transitioning into vortex sectors has been also demonstrated starting from a wave-function obtained from quantum state tomography. Here we are able to reconstruct an exact wave function from single-shot measurements and use this method to ``detect'' the vortex state of the system. This is useful for applications in topological quantum computation where one would need to  ``decode'' input data, then perform a quantum gate operation, and finally create an output. 

Of course, in the case of the Honeycomb model, we have an exact solution of the spectrum using Kitaev's Majorana fermion representation. However, such a representation, though giving exact values for the energies of given states in the Hilbert space, is at the same time losing information. For example, it is difficult and in some cases unknown how to compute expectation values of string operators composed of Pauli matrices. Moreover, one may lose the Majorana representation in cases where a complicated perturbation is added to the Hamiltonian which obscures such a simple representation. For all these reasons it is desirable to obtain realizations of the quantum state in terms of spin eigenstates and the RBM gave in Eq. \eqref{eq:19_0} is such a representation. 

One important application of our results which we present in Section \ref{sec:cft}, is to relate our RBM groundstate wave-function to the Moore-Read Pfaffian state of the Ising CFT. This relation is part of the broader framework of correspondence between 2+1 dimensional topological field theories and 2d CFTs which goes back to the work of Witten on WZW models and Chern-Simons theory \cite{Witten:1988hf,Witten:1991mm}. Our relation is a proposal which passes first non-trivial tests but should be examined in more detail to find further evidence and/or possible proof for it. We leave this for future work. Here, we note that such a correspondence is of great significance as it opens a door for neural networks to represent quantum field theory correlation functions of a high degree of complexity. On the one hand, the parameters of the RBM wave-function grow only polynomially with the spin number, while the CFT correlation function to which it is identified is of exponential complexity in the spin number. We hope that in the future one can generalize this framework/recipe to other models. It would be also very interesting to see whether there is a connection between what we propose, namely using an RBM to compute CFT conformal blocks, and the proposal made in \cite{halverson2020neural} regarding a novel relation between neural networks and quantum field theory.

\phantom{ }

\vfill\break

\begin{acknowledgments}
The work of MN is partially supported by Trinity-Henry Barlow Scholarship. MN would like to thank Yau Mathematical Science Center (YMSC) for hospitality during the early stages of this work. He also thanks Martin Duy Tat for helpful discussions. The work of BH is supported by the National Thousand-Young-Talents Program of China. BH would also like to thank the Bethe Center for Theoretical Physics in Bonn, where part of this work was completed, for hospitality. We would furthermore like to thank Jin Chen, Bartek Czech and Ce Shen for valuable discussions. Part of the numerical calculations were performed on the SARMAD computing cluster at Shahid Beheshti University but the majority of the work was done using the Lambda cluster at YMSC.
\end{acknowledgments}

\section*{Data availability}

The data that support the findings of this study are available from the corresponding author upon reasonable request.


\providecommand{\noopsort}[1]{}\providecommand{\singleletter}[1]{#1}%

\appendix

\onecolumngrid
\section{Symmetries of the Pfaffian Ansatz} \label{sec:PFapp}

\begin{figure}[!ht]
	\centering
	\includegraphics[width=0.8\textwidth]{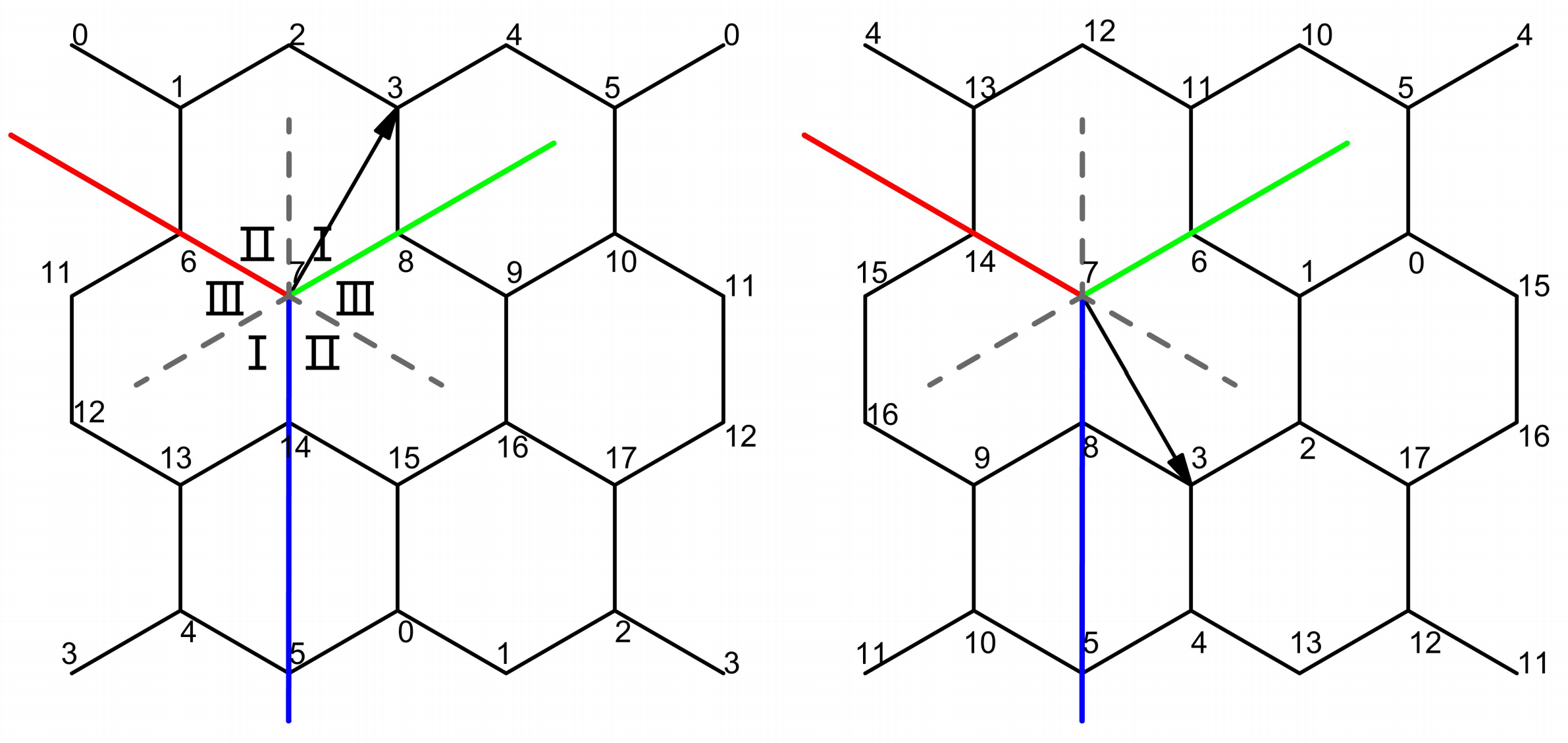}
	\caption{The $\hat{C}_6$ operator for $3\times3$ lattice.}
	\label{fig:c6}
\end{figure}

One remarkable property of the Ansatz \eqref{eq:ansatz} is that the wave-function is invariant under the symmetries of the Honeycomb lattice as well as under the operator which permutes the Pauli matrices. The latter is rather easy to see. Following \cite{Lee_2019}, one can define the operator $\hat{U}_{C_6}$ acting on all lattice sites simultaneously
\begin{equation}
    \hat{U}_{C_6} = \bigotimes_i \frac{1}{2}(\sigma^0_i + i \sigma^x_i + i \sigma^y_i + i \sigma^z_i), 
\end{equation}
and permutes the Pauli matrices as follows
\begin{equation}
    \hat{U}_{C_6} \sigma^x \hat{U}^{\dagger}_{C_6} = \sigma^z, \quad \hat{U}_{C_6} \sigma^y \hat{U}^{\dagger}_{C_6} = \sigma^x, \quad \hat{U}_{C_6} \sigma^z \hat{U}^{\dagger}_{C_6} = \sigma^y.
\end{equation}
Now it is easy to see that $\hat{U}_{C_6}$ leaves our spin-pair states invariant, i.e.
\begin{equation}
    \hat{U}_{C_6} \left(|\uparrow_i\rangle |\downarrow_j\rangle + |\downarrow_i\rangle |\uparrow_j \rangle \right) = |\uparrow_i\rangle |\downarrow_j\rangle + |\downarrow_i\rangle |\uparrow_j \rangle.
\end{equation}
Next, let us show invariance under the $\hat{C}_6$ operator by which we denote the operator generating the 120 degree rotational lattice symmetry. In practice, this operator leads to a re-indexing of sites as depicted in Fig. \ref{fig:c6}. 
Let us now see how our Ansatz is invariant under the action of this operator. To this end, note that $\hat{C}_6$ acts on differences of site coordinates isotropically, namely by a 120 degree rotation. This can be summarized by the following formula
\begin{equation}
    \hat{C}_6 (z_i - z_j) = e^{2\pi i/3} (z_i-z_j). 
\end{equation}
Now note, since $\tau = e^{i \pi/3}$, that this can be rewritten in terms of the complex structure of the torus as follows
\begin{equation}
    \hat{C}_6 (z_i - z_j) = (\tau-1) (z_i-z_j).
\end{equation}
We can then use the transformation properties of theta functions as reviewed in Appendix \ref{sec:theta} to derive the following identity
\begin{eqnarray}
    \vartheta_1(\tau,(\tau-1)(z_i-z_j)) & = & \vartheta_1\left[1/2\atop 1/2\right](\tau,(\tau-1)(z_i-z_j)) \nonumber \\
    ~ & = & \kappa\left(\left[1/2\atop 1/2\right], \gamma\right)^{-1}  (\tau-1)^{-\frac{1}{2}} e^{-i \pi \frac{(z_i-z_j)^2}{\tau-1}} \vartheta_1\left(-\frac{1}{\tau-1},z_i-z_j\right) \nonumber \\
    ~ & = & \kappa\left(\left[1/2\atop 1/2\right], \gamma\right)^{-1}  (\tau-1)^{-\frac{1}{2}} e^{-i \pi \frac{(z_i-z_j)^2}{\tau-1}} \vartheta_1(\tau,z_i-z_j), \label{eq:th1trf}
\end{eqnarray}
where in the first line we have applied the $SL(2,\mathbb{Z})$ transformation
\begin{equation}
    \gamma = \left(\begin{array}{cc}
       0 & -1  \\
       1 & -1
    \end{array}\right),
\end{equation}
and in the last line we have made use of the identity $\frac{-1}{\tau-1} = \tau$ for our particular choice of $\tau$ corresponding to the honeycomb lattice. Similarly, one can derive the identity
\begin{eqnarray}
    \vartheta[\alpha](\tau,(\tau-1)(z_i-z_j) & = & \vartheta\left[\epsilon\atop \epsilon'\right](\tau,(\tau-1)(z_i-z_j)) \nonumber \\
    ~ & = & \kappa\left([\alpha],\gamma\right)^{-1} (\tau-1)^{-\frac{1}{2}}e^{-i \pi \frac{(z_i-z_j)^2}{\tau-1}} \vartheta\left[-\epsilon - \epsilon'-1/2\atop \epsilon\right]\left(-\frac{1}{\tau-1},z_i - z_j\right) \nonumber \\ ~ & = & \kappa\left([\alpha],\gamma\right)^{-1} (\tau-1)^{-\frac{1}{2}}e^{-i \pi \frac{(z_i-z_j)^2}{\tau-1}} \vartheta\left[-\epsilon - \epsilon'-1/2\atop \epsilon\right](\tau,z_i - z_j). \label{eq:thalphatrf}
\end{eqnarray}
Combining equation \eqref{eq:th1trf} and \eqref{eq:thalphatrf}, we then see that the ratio of the two transforms as follows under the action of the $\hat{C}_6$ operator
\begin{equation}
    \hat{C}_6 \frac{\vartheta[\alpha](\tau,z_i-z_j)}{\vartheta_1(\tau,z_i-z_j)} = \frac{\vartheta[\alpha](\tau,(\tau-1)(z_i-z_j))}{\vartheta_1(\tau,(\tau-1)(z_i-z_j))} = \kappa([\alpha],\gamma)^{-1} \kappa\left(\left[1/2\atop 1/2\right], \gamma\right) \frac{\vartheta[\beta](\tau,z_i-z_j)}{\vartheta_1(\tau,z_i-z_j)},
\end{equation}
where the resulting spin structure $\beta$ of the theta function after the transformation is given as follows
\begin{equation}
    \alpha=\left[0\atop 0\right] \longrightarrow \beta = \left[1/2\atop 0\right], \quad \alpha=\left[1/2\atop 0\right] \longrightarrow \beta = \left[0\atop 1/2\right], \quad \alpha = \left[0\atop 1/2\right] \longrightarrow \beta = \left[0 \atop 0\right].
\end{equation}
Thus we see, that as the two factors of $\kappa$ just contribute an overall phase which is the same for all summands of the Pfaffian in expression \eqref{eq:ansatz}, the three groundstate wave functions just get permuted under the action of $\hat{C}_6$.

This is, however, a too large symmetry group as the groundstate space should be only invariant under the \textit{combined} action of $\hat{C}_6$ and $\hat{U}_{C_6}$. To understand what has been happening and how to modify the Ansatz such that it is only invariant under the combined symmetry, it is useful to have another look at the projection \eqref{eq:spinprojection}. When performing such projections, there is a freedom for the choice of phase factors and it is a-priory not clear which ones to choose. In \eqref{eq:spinprojection} we have made the simplest choice for such factors. Here, we want to modify our choice and present a more precise conjecture for the relation between the Pfaffian state and the groundstate wave-functions. To this end, note that the Honeycomb lattice can be divided into three conical regions as shown in Fig. \ref{fig:c6}.
These regions are connected by 120 degree rotations and invariant under inversions. When applying $\hat{P}_{\textrm{physical}}$, we introduce phases depending on in which region the complex value $x_i-z_j$ lies:
\begin{align}
    \widehat{P}_{\textrm{physical}} |ij\rangle_1 & = |\uparrow_i\rangle |\downarrow_j\rangle + |\downarrow_i\rangle |\uparrow_j\rangle \quad & \textrm{ if } z_i - z_j & \in \textrm{region I}~,\\
    \widehat{P}_{\textrm{physical}} |ij\rangle_1 & = |\uparrow_i\rangle |\uparrow_j\rangle  \quad & \textrm{ if } z_i - z_j & \in \textrm{region II}~,\\
    \widehat{P}_{\textrm{physical}} |ij\rangle_1 & = e^{2\pi i/3} |\downarrow_i\rangle |\downarrow_j\rangle \quad & \textrm{ if } z_i - z_j & \in \textrm{region III}.
\end{align}
Next, note the above three spin states are each eigenvectors of $\hat{U}_{C_6}$ with eigenvalues given as follows
\begin{align}
    \hat{U}_{C_6} (|\uparrow_i\rangle |\downarrow_j\rangle + |\downarrow_i\rangle |\uparrow_j\rangle) & = |\uparrow_i\rangle |\downarrow_j\rangle + |\downarrow_i\rangle |\uparrow_j\rangle~, \\ \hat{U}_{C_6} |\uparrow_i\rangle |\uparrow_j\rangle & = e^{2\pi i/3} |\uparrow_i\rangle |\uparrow_j\rangle~, \\
    \hat{U}_{C_6} |\downarrow_i\rangle |\downarrow_j\rangle & = e^{-2\pi i/3} |\downarrow_i\rangle |\downarrow_j\rangle.
\end{align}
Now it is easy to see that a permutation which moves $\widehat{P}_{\textrm{physical}} |ij\rangle_1$ from region I to region II is combined with an action of $\hat{U}_{C_6}$ with eigenvalue $1$, a permutation which moves the state from region II to region III will be combined with an eigenvalue $e^{2\pi i/3}$ from the action of $\hat{U}_{C_6}$ resulting in a state in region III, and a permutation moving the state from region III to I will lead to a cancellation of phases upon the action of $\hat{U}_{C_6}$ resulting in a target state with the correct phase. Thus it is apparent that the Pfaffian state \eqref{eq:ansatz} is invariant (up to a permutation of theta function spin structures as described above) under the combined action of $\hat{C}_6$ and $\hat{U}_{C_6}$ while it is not invariant under individual actions of these operators. This mirrors the symmetry properties of the RBM state under the action of these operators exactly. We therefore conjecture the Pfaffian state to be an exact groundstate under the refined action of $\widehat{P}_{\textrm{physical}}$ as described above.

\section{Theta functions} \label{sec:theta}

We define theta functions with \textit{characteristic} $\left[\epsilon\atop \epsilon'\right] \in \mathbb{R}^2$ as follows:
\begin{equation}
    \vartheta\left[\epsilon\atop \epsilon'\right](\tau,z) = \sum_{n \in \mathbb{Z}}\exp 2 \pi i \left\{ \frac{1}{2}(n + \epsilon)^2 + (n+\epsilon)(z+\epsilon') \right\}.
\end{equation}
The classical Jacobi theta functions are related to these by the relations
\begin{align}
    \vartheta_1(\tau,z) & = \vartheta\left[\frac{1}{2}\atop \frac{1}{2}\right](\tau,z), \\
    \vartheta_2(\tau,z) & = \vartheta\left[\frac{1}{2}\atop 0\right](\tau,z), \\
    \vartheta_3(\tau,z) & = \vartheta\left[0\atop 0\right](\tau,z), \\
    \vartheta_4(\tau,z) & = \vartheta\left[0\atop \frac{1}{2}\right](\tau,z).
\end{align}
The theta functions are periodic, up to a multiplicative factor, under torus lattice translations. Namely, for all integers $m$ and $n$ we have
\begin{equation}
    \vartheta\left[\epsilon\atop \epsilon'\right](\tau,z+n + m \tau) = \exp 2\pi i \left\{n \epsilon - m \epsilon' - m z - \frac{m^2}{2}\tau\right\}\vartheta\left[\epsilon\atop \epsilon'\right](\tau,z).
\end{equation}
Under integral shifts of the characteristics, the theta functions pick up a phase according to
\begin{equation}
    \vartheta\left[\epsilon + m\atop \epsilon' +n\right] (\tau,z) = \exp(2\pi i \epsilon n) \vartheta\left[\epsilon\atop \epsilon'\right](\tau,z),
\end{equation}
and under reflections they satisfy
\begin{equation}
    \vartheta\left[-\epsilon\atop -\epsilon'\right](\tau,z) = \vartheta\left[\epsilon\atop \epsilon'\right](\tau,-z). 
\end{equation}
A property of theta functions deeper than the previous identities is the \textit{modular transformation} property. For any characteristic $\chi \in \left[\epsilon\atop \epsilon'\right] \in \mathbb{R}^2$, and any element $\gamma = \left(\begin{array}{cc}
    a & b  \\
    c & c
\end{array}\right)$ of $SL(2,\mathbb{Z})$, we have
\begin{eqnarray}
    \exp \pi i \left(\frac{-c z^2}{c\tau+d}\right) \vartheta\left[\epsilon\atop \epsilon'\right]\left(\frac{a\tau+b}{c\tau+d},\frac{z}{c\tau+d}\right) = \kappa\left(\left[\epsilon\atop \epsilon'\right], \gamma\right)(c\tau+d)^{\frac{1}{2}} \vartheta\left[a\epsilon+c \epsilon' - \frac{ac}{2}\atop b\epsilon + d \epsilon' + \frac{bd}{2}\right](\tau,z),
\end{eqnarray}
for $z \in \mathbb{C}$, $\tau \in \mathbb{H}^2$, where $\kappa\left(\left[\epsilon\atop \epsilon'\right], \gamma\right)$ satisfies the identity
\begin{equation}
    \kappa\left(\left[\epsilon\atop \epsilon'\right], \gamma \right) = \exp 2\pi i \left\{-\frac{1}{2}(a\epsilon + c \epsilon') bd - \frac{1}{2}(ab \epsilon^2 + cd \epsilon'^2 + 2 b c \epsilon \epsilon') \right\} \kappa\left(\left[0\atop 0\right], \gamma \right),
\end{equation}
with $\kappa\left(\left[0\atop 0\right], \gamma \right)$ an eighth root of unity.


\section{Extended Data}\label{sec8}

\begin{table*}[!htb]
	\centering
	{\rowcolors{2}{green!80!yellow!50}{green!60!yellow!30}
    \begin{tabular}{c|c c c}
    $\alpha$&Number of Hidden Nodes&Number of Parameters&Energy\\
    \hline 
    0.1&  5& 305&-27.7808(59)\\
    0.3& 15& 815&-33.4312(37)\\
    0.5& 25&1325&-36.6024(22)\\
    1.0& 50&2600&-37.2464(13)\\
    1.5& 75&3875&-37.1235(15)\\
    2.0&100&5150&-37.3054(14)\\
    3.0&150&7700&-37.3037(14)\\
    4.0&200&10250&-37.1007(16)\\
    6.0&300&15350&-37.2989(14)\\
    8.0&400&20450&-37.1710(15)\\
    10.0&500&25550&-37.1946(15)\\
    \end{tabular}}
    \vspace{2mm}
	\caption{\label{tab:alpha} Numerical results for the density of hidden nodes, $\alpha = \frac{num_{h}}{num_{v}}$, and its influence on groundstate energy calculation. The results summarized in this table are for a $5 \times 5$ plaquette lattice.} 
\end{table*}

\begin{figure*}[!htb]
	\centering
	\includegraphics[width=\textwidth,height=.4\textheight]{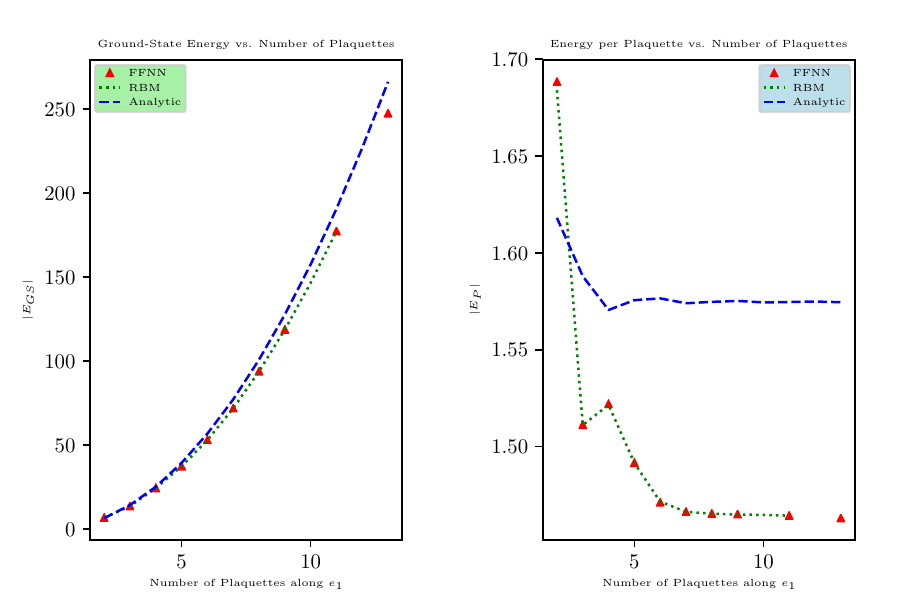}
	\caption{\label{tab:r_space} Numerical results for groundstate energy calculation using Restricted Boltzmann Machines (RBM) 
	and Feed-Forward Neural Networks (FFNN) compared to the analytically obtained exact groundstate energy using Eq.  \eqref{eq:16}.} 
\end{figure*}

\begin{table*}[!htb]
	\centering
    {\rowcolors{2}{green!80!yellow!50}{green!60!yellow!30}
    \begin{tabular}{c|c c c c c c}
    $\alpha$&\# Hidden Nodes&\# Parameters&Energy(FFNN)&Energy(novb-RBM)&Time(FFNN)&Time(novb-RBM)\\
    \hline 
    1.0& 50&2550&-37.2401(15)&-37.1532(15)&2783s&2754s\\
    2.0&100&5100&-37.2958(14)&-36.8954(18)&4591s&5476s\\
    3.0&150&7650&-37.3007(14)&-37.2191(15)&6889s&8331s\\
    \end{tabular}}
    \vspace{2mm}
	\caption{\label{tab:ffnn} Comparison between a single-layer FFNN and a novb-RBM performance, in terms of training time and accuracy of energy for a $5 \times 5$ plaquette lattice.} 
\end{table*}

\end{document}